\documentclass[a4paper,aps,showpacs,10pt]{article}
\pdfoutput=1 

\usepackage{jcappub} 

\usepackage[T1]{fontenc} 

\usepackage{subfig}
\title{\boldmath Loop corrections to primordial fluctuations from inflationary phase transitions }

\author[a]{Yi-Peng Wu,}
\author[a,b,c]{and Jun'ichi Yokoyama}

\affiliation[a]{Research Center for the Early Universe (RESCEU),\\
	Graduate School of Science, The University of Tokyo,
	Tokyo 113-0033, Japan}
\affiliation[b]{Department of Physics,
	Graduate School of Science, \\ The University of Tokyo,
	Tokyo 113-0033, Japan}
\affiliation[c]{Kavli Institute for the Physics and Mathematics of the 
	universe (Kavli IPMU), \\WPI, UTIAS,The University of Tokyo, 5-1-5 Kashiwanoha, Kashiwa 277-8583, Japan}

\emailAdd{ypwu@resceu.s.u-tokyo.ac.jp}
\emailAdd{yokoyama@resceu.s.u-tokyo.ac.jp}

\abstract{We investigate loop corrections to the primordial fluctuations in the single-field inflationary paradigm
from spectator fields that experience a smooth transition of their vacuum expectation values. 
We show that when the phase transition involves a classical evolution effectively driven by a negative mass term from the potential,
important corrections to the curvature perturbation can be generated by field perturbations that are frozen outside the horizon by
the time of the 
phase transition
, yet the correction to tensor perturbation is naturally suppressed by the spatial derivative couplings between
spectator fields and graviton. 
At one-loop level, the dominant channel for the production of primordial fluctuations comes from a pair-scattering of free spectator fields
	that decay into the curvature perturbations, and this decay process is only sensitive
 to field masses comparable to the Hubble scale of inflation.
}


\arxivnumber{1704.05026}

\begin{document} 
\maketitle
\flushbottom

\section{Introduction}
Cosmological correlations in the early universe resolved from the cosmic microwave background and large-scale structure
appear to be Gaussian, adiabatic and nearly scale-invariant \cite{Ade:2015lrj}.
These observed correlations are highly consistent with the natural predictions of inflation  \cite{Starobinsky:1980te,Sato:1980yn,Guth:1980zm,Linde:1981mu,Albrecht:1982wi} (for a review of inflation, see e.g.~\cite{Sato:2015dga,Lyth:1998xn})
	 in the simplest
 scenario where only one single scalar field is relevant.
Albeit at leading order the bilinear average of free fields is sufficient
for distinguishing most of the single-field models, Weinberg \cite{Weinberg:2005vy} argued that
higher-order corrections to the bilinear correlations from nonlinear interactions of the theory may
be as important as the search for non-Gaussian features.

	In the simplest single-field inflation scenario
 the leading (or tree-graph) contributions to Gaussian or non-Gaussian correlations 
are generated around the time where perturbations exit the horizon 
\footnote{
		In multi-field models of inflation, the main contribution to non-Gaussian correlators usually come from the dynamics on superhorizon scales, see for example \cite{Adshead:2008gk,Gao:2009fx}.
}.  However, whether higher-order (loop) corrections are naturally suppressed by the
same assumption remains unclear, and the consequence of loop corrections to inflationary
observables has received much attention
 \cite{Sloth:2006az,Sloth:2006nu,Seery:2007we,Seery:2007wf,Dimastrogiovanni:2008af,Bartolo:2010bu,Adshead:2008gk,Adshead:2009cb,Senatore:2009cf,Senatore:2012nq,Pimentel:2012tw,Senatore:2012ya,Assassi:2012et,Kahya:2010xh}.
  In the single-field inflationary paradigm the curvature perturbation $\zeta$ approaches as a constant well outside the horizon,
  and the tree-level contribution is conserved due to the intrinsic symmetries of inflation on superhorizon scales
  \cite{Weinberg:2003sw,Lyth:2004gb,Langlois:2005qp,Naruko:2011zk,Salopek:1990jq}.
  Loop corrections from $\zeta$ self-interactions are consequently removed by a proper choice of the coordinate
  \cite{Senatore:2009cf,Senatore:2012nq,Pimentel:2012tw,Senatore:2012ya,Assassi:2012et}.
 
On the other hand, loop corrections contributed by additional light fields can be more revealing, given that
isocurvature field perturbations do not necessarily respect all symmetries in the inflationary background.
At least two kinds of isocurvature effects are considered as potential threats to the adiabatic primordial fluctuations.
The first concern is the accumulation of isocurvature perturbations on superhorizon scales, which turns into a logarithmic enhancement
to the curvature perturbation in terms of the ratio between a suitable infrared size and the wavelength of interest 
\cite{Giddings:2010nc,Byrnes:2010yc}. 
The second worry comes from self-interactions of massless spectator fields, such as quartic self-couplings \cite{Weinberg:2005vy,Kahya:2010xh}.
These self-loop corrections simply result in logarithmic enhancement of the power spectrum in terms of the scale factor during inflation.
While the former effect is subject to the issue on the infrared divergence from isocurvature degrees of freedom \cite{Geshnizjani:2002wp,Geshnizjani:2003cn,Senatore:2012nq,Bartolo:2007ti,Lyth:2007jh} (see \cite{Tanaka:2013caa} for a review),
the latter effect disappears when all possible loop diagrams are taken into account \cite{Senatore:2009cf}.
In this work, we intend to draw out another kind of temporal enhancement led by isocurvature modes running in loops.
(See also \cite{Gao:2009fx,Gao:2009qy,Biagetti:2013kwa,Biagetti:2014asa,Fujita:2014oba,Suyama:2014vga} 
for yet another enhancing mechanism to primordial fluctuations from generalized spectator fields.)

The loop effect of isocurvature fields with non-negligible masses has also been addressed in many previous studies \cite{Weinberg:2006ac,Xue:2012wi,Saito:2012pd,Tanaka:2015aza}.
Indeed, massive fields exhibit interesting behaviors in both classical and quantum regimes. 
Isocurvature modes with masses larger than the size of the Hubble scale are oscillating, creating a time-dependent background for their quantum fluctuations
that may characterize important informations in the primordial Universe \cite{Burgess:2002ub,Chen:2015lza,Liu:2016aaf}.
However, with 
	 a mass term smaller than the Hubble scale
the perturbative expansion of the corrections can be summed up into an effective mode function that
always decays outside the horizon. As a result, higher-order corrections from a massive field converge to a constant 
no much different from the value evaluated around horizon exit \cite{Weinberg:2006ac,Senatore:2009cf}, similar to the case of a massless field.
This generic conclusion shall hold for massive fields that always stay in one stable vacuum state during inflation.

In this work, we take into account the possibility that some fields may experience a transition of their vacuum expectation values during inflation.
The transition can happen instantly due to quantum tunneling or jumping, or it can take place smoothly under a classical evolution.
Both kinds of phase transitions are well motivated since they might be the origin of inflation \cite{Starobinsky:1980te,Sato:1980yn,Guth:1980zm,Linde:1981mu,Albrecht:1982wi}.
In particular during a smooth transition of the field values one usually finds a period that the relevant field has 
a negative effective mass square (see for example \cite{Nagasawa:1991zr}).
Although these fields end up in some stable vacua at most of the times, the phase with negative mass square can lead to the growth of field
perturbations even on superhorizon scales, and thus, also lead to the growth of loop corrections. 
The aim of this study is to clarify how large those growing corrections may reside in
the cosmological observables.

Our calculations must focus on contributions from superhorizon modes of isocurvature perturbations, 
given that a mass term does not change the behavior of a field with sufficiently large wavenumbers.
For our purpose it turns out to be convenient to adopt a specific perturbative expansion of the loop diagrams proposed by 
Senatore and Zaldarriaga \cite{Senatore:2009cf}, 
based on the modified Feynman rules for cosmological correlation functions developed earlier by Musso \cite{Musso:2006pt}.
	To compute a bilinear correlator in this specific scheme one simply rearrange Weinberg's formalism \cite{Weinberg:2005vy} to double the number of
nested commutators, whereas the commutator form is most convenient for reading informations in the IR (superhorizon) regime \cite{Chen:2010xka}.
Following the arrangement of the loop diagrams in \cite{Pimentel:2012tw,Senatore:2009cf} 
we clearly identify the dominant contribution from the mass-induced interactions 
in the late-time limit where all perturbative fields are sufficiently classicalized.

This article is organized as follows. In section~\ref{sec:II}, we review loop corrections from massive fields with an extended discussion on the case
of negative effective masses. We show the existence of a negative mass term in the transition phase of field expectation values by a sampling model
given in section~\ref{sec:III}, and in section~\ref{sec:IV} we calculate the one-loop correction to the curvature perturbation and the tensor
perturbation based on the given sample. We discuss the possible contribution from higher-order perturbations beyond one-loop 
in section~\ref{sec:higher-order} and, finally, our conclusion is summarized in section~\ref{sec:conclusion}.

\section{Loop corrections from massive fields}\label{sec:II}
In this section we investigate loop corrections to the curvature perturbation induced by spectator fields only through gravitational interactions.
For massless fields any interaction in the Lagrangian must involve at least one spatial or temporal derivative
so that their contribution are suppressed on scales well outside the horizon.
As a result the curvature perturbation contributed by massless fields at loop level is always frozen sometime after horizon crossing \cite{Senatore:2009cf}.
The situation beocmes more interesting when a spectator field gains interactions without derivatives from a mass term.
Higher-order interactions with no derivative can result in time-evolving corrections on superhorizon scales, essentially sourced 
by the evolution of the wavefunction of a massive field.  
However, these loop corrections always decay with a positive mass term, as discussed in \cite{Weinberg:2006ac,Senatore:2009cf}.
Here we consider a simple example to see the time-dependence of the induced curvature perturbation,
but treating the mass square as a free parameter with the possibilty to be negative.  

\paragraph{The inflationary paradigm.}
To make a concrete discussion, we introduce two real scalars, namely, 
the inflaton field $\phi(\mathbf{x}, t)$ whose quantum fluctuations generate the large-scale
curvature perturbations $\zeta(\mathbf{x}, t)$ at the time of horizon exit, and a spectator field 
$\sigma(\mathbf{x}, t)$ with insignificant energy density during inflation. 
The action of our interest is therefore given by
\begin{equation}\label{model}
S=\int d^4 x \sqrt{-g} \left[ \frac{M_p^2}{2}R  -\frac{1}{2}(\partial \phi)^2 -\frac{1}{2}(\partial\sigma)^2 - V(\phi, \sigma) \right] ,
\end{equation}
where $M_p$ is the reduced Planck mass, and we shall specify the potential in the next section.

We apply the metric of the Arnowitt-Deser-Minser (ADM) formalism 
\begin{equation}
\label{eq:ADM_t}
ds^2 = - N^2 dt^2 + \gamma_{ij}(dx^i +  N^i dt)(dx^j + N^j dt),
\end{equation}
and take the parametrization in the spatially flat slicing to the lapse function, the shift function and the spatial metric as
\begin{equation}
\label{def:ADM_parametrize}
\alpha = N-1, \;\;\; \beta^i = a N^i , \;\;\; \gamma_{ij}= a^2(e^h)_{ij},
\end{equation}
where $\beta^i = \partial_i \beta + \beta^i_T$, $\partial_i\beta^i_T =0 $ and $h_{ii}=\partial_ih_{ij}=0$.
The ADM metric in terms of the conformal time $\eta = -\int_{t}^{\infty} dt/a$ thus reads
\begin{equation}
\label{eq:metric_eta}
ds^2 = -a^2(\eta) [(1+\alpha)^2 d\eta^2 + (e^h)_{ij}(dx^i +  \beta^i d\eta)(dx^j + \beta^j d\eta)].
\end{equation}
The scale factor during inflation $a(\eta)\sim -1/(H \eta)$ diverges as $\eta \rightarrow 0$, 
where the Hubble parameter $H = \mathcal{H}/a$ is nearly a constant.
Let us decompose the inflaton field into homogeneous parts and pertubations as $\phi(\mathbf{x}, \eta) =\phi(\eta) + \delta\phi(\mathbf{x}, \eta)$.
Following the computation in \cite{Maldacena:2002vr} and omitting any contribution from $\sigma$, we obtain the solutions
for linear perturbations as (see appendix \ref{Appendix} for more details)
\begin{equation}
\label{sol:flat slicing}
\alpha\simeq -\epsilon_\phi \zeta,\;\;\; \partial^2\beta \simeq \epsilon_\phi \zeta^\prime, \;\;\; 
\delta\phi\simeq - \frac{\phi^\prime}{\mathcal{H}}\zeta = -\frac{\sqrt{2\epsilon_\phi}}{\kappa}\zeta,
\end{equation}
where $\epsilon_\phi = \kappa^2\phi^{\prime \, 2}/(2 \mathcal{H}^2)$ and $\phi^\prime\equiv d\phi/d\eta$.
Inserting the solutions \eqref{sol:flat slicing} back into our theory \eqref{model}, we recover the quadratic 
 action for the curvature perturbation in the single field inflationary scenario
\begin{equation}\label{eq:tree zeta}
S^{(2)}_\zeta =\int d\eta d^{3}x\; \mathcal{L}^{\rm tree}_{\zeta}
= \int d\eta d^{3}x\; \frac{\epsilon_\phi}{\kappa^2} \,a^{2}\left[ (\zeta^{\prime })^2 -(\partial\zeta)^2\right].
\end{equation}
The free-field operator $(2\pi)^{3}\zeta_I(\mathbf{x},\eta) = \int d^3k \, e^{i\mathbf{k}\cdot\mathbf{x}}\zeta_{\mathbf{k}}(\eta)$
is then defined from the solution of a field equation derived from \eqref{eq:tree zeta}, where
	 the subscript $I$ stands for fields in the interaction picture and
the mode function $\zeta_{\mathbf{k}} \equiv \zeta_k\hat{a}_\mathbf{k}+\zeta_k^\ast\hat{a}^\dagger_\mathbf{-k}$ 
normalized by the usual vacuum in the Minkowski limit is well-known:
\begin{equation}
\label{eq:mode_zeta}
\zeta_k=\frac{\kappa H}{\sqrt{2\epsilon_\phi}\sqrt{2k^3}}e^{-i k\eta}(i - k\eta).
\end{equation}

\paragraph{The massive mode.}
We now consider the spectator field $\sigma$ that has only gravitational interaction with $\phi$, which is manifest by the condition
$\partial^2V/(\partial\phi\partial\sigma)\equiv V_{\phi\sigma}=0$. In the ADM formalism \eqref{eq:ADM_t},
the action of $\sigma$ is given by
\begin{equation}\label{eq:action_sigma general}
S_\sigma=\int d^4 x \sqrt{\gamma} \left[N^{-1}(\dot{\sigma}- N^i\partial_i\sigma)^2 -N\gamma^{ij}\partial_i\sigma\partial_j\sigma - 2NV\right],
\end{equation}
	 where $\gamma \equiv \mathrm{det}(\gamma_{ij})$ with $\gamma_{ij}$ given by \eqref{def:ADM_parametrize}.
For convenience, we consider that the spectator field has a well-defined decomposition as 
$\sigma(\mathbf{x},\eta) =\sigma(\eta) + \delta\sigma(\mathbf{x}, \eta)$, where the classical equation of motion for the homogeneous part $\sigma(\eta)$
is given in appendix \ref{Appendix}. We will treat the field potential as a non-perturbative contribution to the free field solutions.
In the Fourier space the equation of motion for the field perturbation is then reads
\begin{equation}\label{eq:eom_free sigma_k}
\delta\sigma^{\prime\prime}_{\mathbf{k}}+(2\nu - 1)\mathcal{H}\delta\sigma^{\prime}_{\mathbf{k}}
+ (k^2 + a^2 V_{\sigma\sigma})\delta\sigma_{\mathbf{k}} =0.
\end{equation}
Note that we have promoted the theory into $2\nu$ space dimensions
as considered in \cite{Weinberg:2005vy}, where $\nu \rightarrow 3/2$ recovers the usual space dimensionality 3.
Assuming that $M^2\equiv a^2V_{\sigma\sigma}/\mathcal{H}^2 = V_{\sigma\sigma}/H^2$ is a constant, 
the mode function is solved as
\begin{equation}
\label{eq:mode_sigma}
\delta\sigma_k(\eta)= \frac{\sqrt{\pi}}{2} e^{i(L+\frac{1}{2})\frac{\pi}{2}}H^{\nu-\frac{1}{2}}\left(-\eta\right)^\nu H^{(1)}_L(- k \eta), \;\;\; L =\sqrt{\nu^2 -M^2}.
\end{equation}
where $H^{(1)}_L$ is the Hankel function of the first kind with $L>0$.
The free-field operator is then given by 
$(2\pi)^{3}\sigma_I(\mathbf{x},\eta) = \int d^3k \, e^{i\mathbf{k}\cdot\mathbf{x}}\sigma_{\mathbf{k}}(\eta)$,
where $\sigma_{\mathbf{k}} \equiv \sigma_k\hat{a}_\mathbf{k}+\sigma_k^\ast\hat{a}^\dagger_\mathbf{-k}$. 
  For imaginary $L$ led by $M > \nu$ the perturbation $\delta\sigma_k$
vanishes soon after horizon exit \cite{Weinberg:2006ac}. Expanding the mode function \eqref{eq:mode_sigma} 
around $\vert k\eta\vert \rightarrow 0$, we get
\begin{align}
\label{eq:mode_sigma_latetime}
\delta\sigma_k(\eta)= \frac{\sqrt{\pi}}{2} e^{i(L+\frac{1}{2})\frac{\pi}{2}}H^{\nu-\frac{1}{2}} &\left(-\eta\right)^\nu
\left[(-k\eta)^L \frac{2^{-L}}{\Gamma(L+1)} \right. \nonumber\\
&\left. -\frac{i}{\pi}2^L\Gamma(L) (-k\eta)^{-L}+ \cdots \right].
\end{align}
Since we consider $L > 0$, the term in the second line of \eqref{eq:mode_sigma_latetime} dominates as $\vert k\eta\vert \rightarrow 0$,
so that $\delta\sigma_k\sim a^{L-\nu}$ at late times. 
On the other hand we can obtain the late-time behavior of the (retarded) Green function 
\begin{align}
\label{eq:Green_sigma}
G_\sigma(x,x_1) &=i\Theta(\eta - \eta_1)[\sigma(\mathbf{x},\eta),\sigma(\mathbf{x}_1, \eta_1)], \\
&= i\Theta(\eta - \eta_1) \int \frac{d^3 \mathbf{k}}{(2\pi)^3} e^{i\mathbf{k}(\mathbf{x}-\mathbf{x}_1)}
\left[\delta\sigma_k(\eta)\delta\sigma^\ast_k(\eta_1)-\delta\sigma_k(\eta_1)\delta\sigma^\ast_k(\eta)\right],
\end{align}
from the expansion \eqref{eq:mode_sigma_latetime}. To the lowest order in $\vert k\eta\vert$ with the condition
$\eta_1 \leq \eta\leq 0$, one finds $G_\sigma \sim a(\eta)^{L-\nu} a(\eta_1)^{-L-\nu}$ as $\eta\rightarrow 0$.
	The Green function for $\zeta$, namely $G_\zeta (x,x_1)$, can be computed from a similar way as \eqref{eq:Green_sigma}. If we ignore the corrections from the slow-roll parameter, the late-time behavior is estimated as $G_\zeta\sim a(\eta_1)^{-2\nu}$ when taking $\eta\rightarrow 0$.


As shown in \cite{Musso:2006pt}, the free-field operators $\zeta_I$, $\sigma_I$ in the interaction picture and the Green functions $G_\zeta$, $G_\sigma$
are the only ingredients that we need to compute correlation functions to all loop levels.
For instance, in $2\nu$ space dimensions the cubic interactions between $\zeta$ and $\sigma$ given by \eqref{eq:action_sigma general}  are 
\begin{align}\label{eq:cubic terms}
S^{(3)}_\sigma = \int d\eta d^{2\nu}x &\frac{a^{2\nu -1}}{2} \left[-\alpha \delta\sigma^{\prime 2} -\alpha(\partial\delta\sigma)^2-2\delta\sigma^\prime(\partial^i\beta\partial_i\delta\sigma) 
\right. \\\nonumber
&\left.  
+2\sigma^\prime\alpha^2\delta\sigma^\prime -\sigma^{\prime\, 2}\alpha^3 + 2\sigma^\prime\alpha (\partial^i\beta\partial_i\delta\sigma)-a^2\alpha(V_{\sigma\sigma}\delta\sigma^2)
\right],
\end{align}
where the inflaton perturbation $\delta\phi$ and the metric pertubations $\alpha$, $\beta$ are to be
replaced by $\zeta_I$ according to \eqref{sol:flat slicing}.
As mentioned, the interaction without field derivatives has the highest powers of $a$, which we denoted as
\begin{align}
	\label{def:nonderivative_interaction}
\mathcal{L}_1 = \epsilon_\phi\frac{a^{2\nu+1}}{2}   V_{\sigma\sigma} \zeta\delta\sigma^2.
\end{align}
For $\vert k\eta\vert \rightarrow 0$, the dominant one-loop correction to the curvature perturbation induced
by $\mathcal{L}_1$ can be schematically given by the Green's function method (we will come back to this point in Sec. \ref{sec:IV}, see also \cite{Senatore:2009cf})
\begin{align}
\zeta_1(\mathbf{x},\eta)=-\frac{V_{\sigma\sigma}}{2}\int d^3x_1\int^\eta d\eta_1 \left[\sqrt{2\epsilon_\phi}a^2(\eta_1)\right]^{\nu + \frac{1}{2}}
G_\zeta(\mathbf{x},\mathbf{x}_1) \delta\sigma^2(\mathbf{x}_1,\eta_1).
\end{align}
Taking the above approximation for $G_\zeta$ and $\delta\sigma$, we estimate the late-time behavior of the induced
curvature perturbation as
\begin{equation}\label{eq:zeta1_latetime}
\zeta_1 \rightarrow \int^\eta d\eta_1\;\eta_1^{2\nu-2L-1} \sim
 \left\lbrace 
\begin{array}{ll}
a^{2L - 2\nu}, \;\;&\; L\neq\nu, \\
\ln a, \;\;&\; L =\nu.
\end{array} \right.
\end{equation}
Therefore with $M^2 > 0$, we have $L < \nu$ such that $\zeta_1$ decays on superhorizon scales.
In this case $\zeta_1$ is dominated by the value evaluated around the time of horizon exit, and the contribution from all the other
derivative interactions shown in \eqref{eq:cubic terms} should be taken into account.   
In the massless limit ($M\rightarrow 0$) the correction $\zeta_1$ vanishes as $V_{\sigma\sigma}\rightarrow 0$.

We remark that the interesting behavior can appear for a massive mode with $M^2 < 0$, where $L > \nu$ and $\zeta_1\sim a^{2L - 2\nu}$
is growing with time. Ignoring for the moment the issue of negative mass states in field theory,
we emphasize that the time evolution of $\zeta_1$ on superhorizon is physical and is not cancelled by other loop diagrams, 
as the corrections induced by massive modes with positive mass square.


\section{Inflationary phase transitions}\label{sec:III}
The nature of inflation remains unknown, and it is believed that there exist a large number of metastable vacua
around the energy scale of inflation such that fields may transit from one local minimum to another.
The transition could occur instantly through quantum tunneling or jumping, which is subject to first-order phase transitions.
On the other hand, the transition could be smooth if one field moves to a new vacuum by (slow-)rolling.
The latter case is refered to as second-order phase transitions which generically exhibit well-defined coherent 
dynamics over the horizon scale and will be the main focus of this work.

The potential $V(\phi,\sigma)$ given in \eqref{model} 
is at least composed by an inflaton potential $V_0(\phi)$ and a subdominant potential $V_1(\sigma)$ of the field $\sigma$ as
\begin{equation}
V(\phi, \sigma) = V_0(\phi) + V_1(\sigma) + \dots,
\end{equation}
where the background density is attributed to $V_0(\phi)\approx 3 M_p^2 H^2$. 
To our purpose it is convenient to consider a quasi de Sitter expansion
charaterized by the slow-roll parameter $\epsilon_H = -\dot{H}/H^2$. Provided $0<\epsilon_H \ll 1$, the Hubble parameter 
$H \sim a^{-\epsilon_H}$ decays slowly with time, and so as the background density $V_0(\phi)$. 

\begin{figure}
	\begin{center}
		\includegraphics[width=6.2cm]{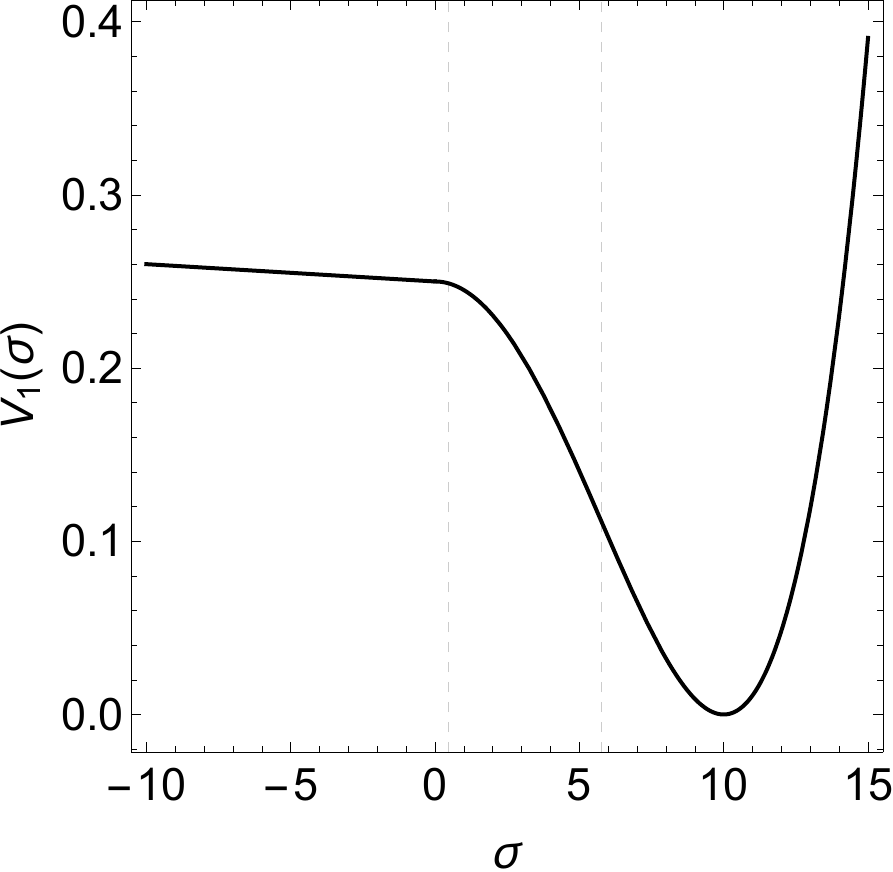}
		\hfill
		\includegraphics[width=6.5cm]{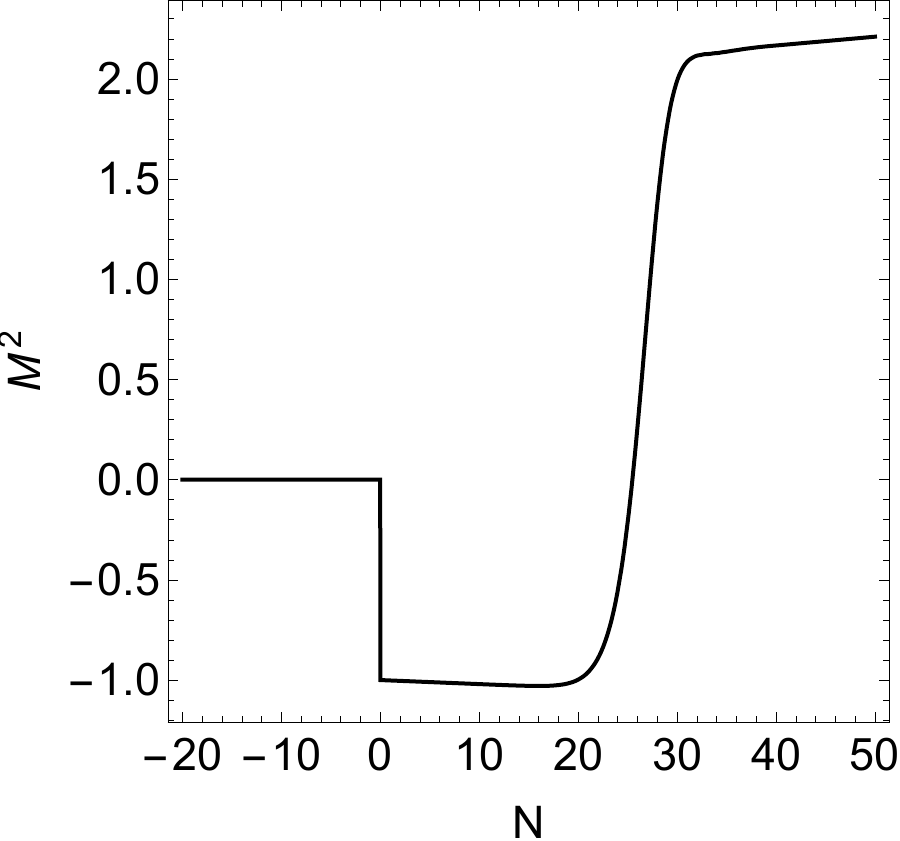}
	\end{center}
	\caption{\label{Fig_1} Evolution of the mass parameter $M^2$ during the classical transition of $\sigma$ in a quasi de Sitter expansion with a fixed
		slow-roll parameter $\epsilon_H =0.001$, where the initial conditions satisfy
		$\sigma = 0$ and $d\sigma/dN = g^3/(3H_i^2)$ at $N= 0$. Both $V_1$ and $\sigma$ are shown in the unit of $H_i = 1\times 10^{-6} M_p$.
	The potential parameters are $\lambda = 0.01$, $v = 10 H_i$, and $g = 0.2 H_i$.
}
\end{figure}

	Many inflationary models exhibit a plateau region in their field potentials for slow-rolling,
and these fields can exit from the phase of slow-roll as they roll down into some local minima.	
The intermediate region that connects the plateau and a local minimum is usually convex upward, which effectively generates a 
negative mass term and speeds up the field evolution.
Typical potentials of this kind can be found in the class of small-field inflation \cite{Lyth:1998xn,Brax:2010ai,Linde:1981mu,Albrecht:1982wi}, 
such as the hilltop inflation \cite{Boubekeur:2005zm}. The inflationary potential that has a convex-upward region (or a tachyonic region)
can also be found in the $R^2$ inflation \cite{Starobinsky:1980te}, the Higgs inflation with a non-minimal coupling \cite{Bezrukov:2007ep} 
and the hybrid inflation \cite{Linde:1993cn}.
However, analytical study of the field dynamics in the tachyonic region appears to be very difficult in those well-motivated models.
For simplicity, let us introduce a specific potential of the form
\begin{align}\label{eq:potential_sigma}
V_1( \sigma) =& -g^3 \sigma + \frac{\lambda}{4}v^4, & \sigma < 0, \\\nonumber
=&  \frac{\lambda}{4}(\sigma^2 - v^2)^2  & \sigma \geq 0,
\end{align}
as seen by Fig. \ref{Fig_1}.
We also introduce a dimensionless mass parameter
\begin{equation}\label{def:M^2}
M^2\equiv V_{\sigma\sigma}/H^2 =  \left\lbrace 
\begin{array}{ll}
0\; , \;\;&\; \sigma < 0, \\
\lambda(3\sigma^2 -v^2)/H^2, \;\;&\; \sigma\geq 0.
\end{array} \right. 
\end{equation}


\paragraph{The case with $g=0$.}
If we assume the initial condition $\sigma < 0$ so that $\sigma$ is in a perfectly flat region at the beginning,
 the behavior of $\sigma$ is governed by its quantum fluctuation with a typical size
$\Delta\sigma_q \sim \frac{H}{2\pi}$. The field value at the next moment has no predictability, but there is a
global minimum at $\sigma = v$. Suppose that $\sigma$ lands around the origin ($\sigma\rightarrow 0^+$) accidentally at a moment $\eta_i$,
it suddenly feels a mass $M^2 \approx -\lambda v^2/H_i^2 < 0$, where $H_i $ is the Hubble parameter at $\eta = \eta_i$.
This negative mass tend to drag $\sigma$ towards the global minimum. 
As long as the classical deviation $\Delta\sigma_c$ dominates the quantum fluctuation $\Delta\sigma_q $, 
the behavior of the $\sigma$ field can be described by
the classical equation of motion \eqref{eq:eom_sigma_classical}, and for $\sigma > 0$ we have
\begin{equation}
\label{eq:eom_sigma_N}
\frac{d^2 \sigma}{d N^2} + (3 - \epsilon_H) \frac{d \sigma}{d N} + \left[ \frac{\lambda}{H^2}(\sigma^2 - v^2)\right]\sigma =0,
\end{equation}
where $N \equiv \ln a$ is the e-folding number.

At some value $\sigma = \sigma_i > 0$, the classical deviation $\Delta\sigma_c$ start to surpass the quantum effect $\Delta\sigma_q$ and
the field will start to roll down the plateau. Since the transition is smooth for $\vert M^2\vert\sim  \mathcal{O}(1)$, 
we neglect the term $d^2\sigma/dN^2$ so that \eqref{eq:eom_sigma_N} becomes
\begin{equation}
\frac{d \sigma}{d N} = \frac{\lambda v^2}{3H^2}\sigma,
\end{equation}
for $\sigma\ll v$, where we can omit the correction from $\epsilon_H$. Taking $H\approx H_i e^{-\epsilon_{i}(N-N_i)}$,
where $\epsilon$ is a constant value of $\epsilon_H$, the solution of \eqref{eq:eom_sigma_N} is given by
\begin{equation}
\label{sol:sigma_slowroll}
\sigma = \sigma_i \exp\left[\frac{1}{2\epsilon}\frac{\lambda v^2}{3H_i^2}\left(e^{2\epsilon (N-N_i)}-1\right)\right].
\end{equation}
The value of $\sigma_i$ can be estimated by using the classical deviation within one e-fold as 
$\Delta\sigma_c \sim \frac{\lambda v^2}{3H^2}\sigma$, and imposing the condition $\Delta\sigma_c \geq  \frac{H}{2\pi}$, to yield
$\sigma_i\geq 3H_i^3/(2\pi \lambda v^2)$.

Once $\sigma$ reaches the value $v/\sqrt{3}$, the mass square $M^2$ starts to change from negative to positive.  
We can estimate the duration of the slow roll in a pure de Sitter expansion by taking $\epsilon \rightarrow 0$,
where \eqref{sol:sigma_slowroll} shows
\begin{equation}
\sigma \rightarrow\sigma_i \exp \left[\frac{\lambda v^2}{3H_i^2}(N-N_i)\right].
\end{equation}
	We estimate the phase of slow roll ends at $\sigma_f \equiv \sigma(N_f)$ by the condition $\vert\sigma V_{1\sigma} / V_1 \vert_{\sigma = \sigma_f} =1$,
	where $\sigma_f < v \ll M_p$. In fact, one can solve the condition to find that $\sigma_f = v/\sqrt{3}$.
	 It is also useful to define the critical mass at $\sigma\gtrsim 0$ from \eqref{eq:potential_sigma} as $M_i^2\equiv \lambda v^2/H_i^2$, then the duration reads
\begin{equation}
\label{eq:duration_of_slowroll}
\Delta N = N_f- N_i = \frac{3}{M_i^2} \ln \left(\frac{2\pi}{3\sqrt{3\lambda}}M_i^{3}\right).
\end{equation}
The condition $\Delta N > 0$ implies $\frac{4\pi^2}{27}M_i^6 > \lambda$.
The result of \eqref{eq:duration_of_slowroll} implies $\Delta N \rightarrow 0 $ as $M_i \rightarrow\infty$,
given that in the large mass limit the rolling finishes immediately as similar to an event of a first-order phase transition.

If the field mass is small compared with the Hubble parameter ($\vert M^2\vert\ll 1$), 
quantum effects from short wavelength fluctuations become important. In this case the behavior of $\sigma$
can be studied by the stochastic approach \cite{Starobinsky:1986,Starobinsky:1994bd}
\begin{equation}\label{eq:Langevin1}
\frac{d \sigma}{d N} = -\frac{1}{3 H^2}\frac{\partial V(\phi, \sigma)}{\partial\sigma} + \frac{f(N)}{H},
\end{equation}
where the quantum noise $f(N)$ satifies the correlation
$\langle f(N_1)f(N_2)\rangle \approx H^4\delta(N_1 -N_2)/(4\pi^2)$.
As the field is slowly rolling, the equation motion of $\langle \sigma^2\rangle$ derived from \eqref{eq:Langevin1} shows 
\begin{equation}
\frac{d}{d N} \langle \sigma^2\rangle - 
\frac{2 \lambda v^2}{3 H^2} \langle \sigma^2\rangle + \frac{2\lambda}{3 H^2} \langle \sigma^4\rangle  =\frac{H^2}{4\pi^2},
\end{equation}
where the term $\langle \sigma^4\rangle$ becomes less important as the slow-roll phase starts from 
$\sigma\rightarrow\sigma_i$ near the origin \cite{Nagasawa:1991zr}.
In a pure de Sitter expansion $H \rightarrow H_i $ we can easily find the solution
\begin{equation}
\langle \sigma^2\rangle = \sigma_i^2 e^{\frac{2}{3}M_i^2 (N-N_i)} + \frac{M_i^2}{6\pi^2}\left(e^{\frac{2}{3}M_i^2 (N-N_i)} -1\right),
\end{equation}
and the duration of slow roll is obtained by using $\sqrt{\langle \sigma^2\rangle } \rightarrow v/\sqrt{3}$, which results in
\begin{equation}
\label{eq:DeltaN_small_mass}
\Delta N =  \frac{3}{2 M_i^2} \ln \left[\frac{M_i^6(2 + 4\pi^2/\lambda)}{27+ 2 M_i^6}\right].
\end{equation}
For $M_i^2 \ll 1$ and $\lambda < 1$ the factor in the logarithmic function
$(2 + 4\pi^2/\lambda)/(27+ 2 M_i^6) \approx  4\pi^2/(27 \lambda)$ so that \eqref{eq:DeltaN_small_mass}
coincides with the result of \eqref{eq:duration_of_slowroll}. 
	Therefore, in the following discussion we are safe to apply \eqref{eq:duration_of_slowroll} for both the regime $M_i^2 \geq 1$ and $M_i^2 < 1$.

\paragraph{The case with $g >0$.}
In general, the plateau	region ($\sigma < 0$) may slightly tilt towards the global minimum such that 
the evolution of $\sigma$ becomes deterministic with time. For example, if $\sigma$ starts to roll down from a value $\sigma_0 < 0$ at $N = N_0$, 
the equation of motion for the linear potential \eqref{eq:potential_sigma} can be approximated as
\begin{equation}\label{eq:eom_sigma_N_g>0}
\frac{d \sigma}{d N} = \frac{g^3}{(3-\epsilon)H^2}e^{2 \epsilon (N-N_0)},
\end{equation}
which leads to the solution of the form
\begin{equation}
\label{sol:sigma_slowroll g>0}
\sigma = \sigma_0 +\frac{1}{2\epsilon}\frac{g^3}{(3-\epsilon)H_i^2} \left[e^{2\epsilon (N-N_0)}-1\right].
\end{equation}

We expect the first phase transition occurs as $\sigma$ crosses the origin, 
where the equation of motion in the region $0< \sigma \ll v$ reads  
\begin{equation}
\label{eq:eom_sigma_N2}
\frac{d^2 \sigma}{d N^2} + (3 - \epsilon) \frac{d \sigma}{d N} - M_i^2 e^{2\epsilon (N-N_i)} \sigma =0.
\end{equation}
Without the loss of generality, we take $\sigma_i \equiv \sigma (N_i) =0 $ and set $N_i = 0$ for convenience.
With the field velocity given by $\sigma^\prime(0) =g^3e^{-2\epsilon N_0}/(3H_i^2)$, the solution of \eqref{eq:eom_sigma_N2} is
\begin{align}
\label{sol:sigma g>0}
\sigma =& \frac{\pi}{2\epsilon}\frac{g^3}{3H_i^2}\csc\left(\frac{3\pi}{2\epsilon}\right) 
e^{-2\epsilon\, N_0}e^{-3N/2} \\\nonumber
& \times \left[I_{-\frac{3}{2\epsilon}}\left(\frac{M_i}{\epsilon}\right) I_{\frac{3}{2\epsilon}}\left(\frac{M_i}{\epsilon}e^{\epsilon\, N}\right)
-I_{\frac{3}{2\epsilon}}\left(\frac{M_i}{\epsilon}\right)I_{-\frac{3}{2\epsilon}}\left(\frac{ M_i}{\epsilon}e^{\epsilon\, N}\right)\right],
\end{align}
where $I_n$ is the modified Bessel function of the first kind.
In the limit of $\epsilon \rightarrow 0$, the solution \eqref{sol:sigma_slowroll g>0} shows 
$\sigma\rightarrow \sigma_0 +g^3(N-N_0)/(3H_i^2)$, and the solution of the equation of motion \eqref{eq:eom_sigma_N2} 
in the second phase is simplified as 
\begin{align}
\sigma \rightarrow \sigma_+ e^{(l-\frac{3}{2})N} + \sigma_- e^{-(l+\frac{3}{2})N},
\end{align}
where $l = \sqrt{9/4 + M_i^2}$ and $\sigma_+ = - \sigma_- = g^3/(6l H_i^2)$.
Since $l > 3/2$, $\sigma_+$ ($\sigma_-$) corresponds to the growing (decaying) mode of the solution.

As the phase with a negative mass (the tachyonic phase) ends at $\sigma(N_f) = \sigma_f = v/\sqrt{3}$, 
one can neglect the contribution from the decaying mode if $e^{-2l N_f}\ll 1$. The duration of the tachyonic phase is then estimated by
\begin{align}\label{eq:DeltaN g>0}
	\Delta N = \frac{1}{l-3/2}\ln\left[\frac{6l v H_i^2}{\sqrt{3}g^3}\right],
\end{align}
	where we can replace the parameter $v$ with $M_i$ by the definition $M_i^2\equiv \lambda v^2/H_i^2$.
The condition $\Delta N > 0$ imposes $2\sqrt{3} l M_i > g^3/H_i^3$.
In the limit $M_i \rightarrow \infty$, we find that $l \rightarrow M_i$ and the duration $\Delta N \rightarrow 0$.

\section{One-loop corrections}\label{sec:IV}
To calculate loop corrections on superhorizon scales it is convenient to adopt the diagrammatic expansion scheme performed in \cite{Senatore:2009cf,Pimentel:2012tw}, based on the modified Feynman rules developed in \cite{Musso:2006pt}.
In this scheme, higher-order perturbations are always resolved into solutions of the free fields by virtue of the Green's method,
and therefore correlation functions are simply resulted from convolutions of free fields with retarded Green functions.
For example, the $\zeta$ two-point function is given by
\begin{align}\nonumber
\langle\Omega\vert \zeta^2(\eta)\vert\Omega\rangle &= \langle 0\vert U_I^\dagger(\eta, -\infty_+)\zeta^2_I(\eta) U_I(\eta, -\infty_+)\vert 0 \rangle, \\
\label{eq:free_zeta evolution}
&= \langle U_I^\dagger(\eta, -\infty_+)\zeta_I(\eta)U_I(\eta, \eta_1) U_I^\dagger(\eta, \eta_1)\zeta_I(\eta)U_I(\eta, -\infty_+) \rangle ,
\end{align}
where $\vert\Omega\rangle$ is the vacuum of the interacting theory and $\vert 0 \rangle$ is the vacuum of the free theory.
The $-\infty_+$ implies a replacement of $\eta_1$ by $\eta_1(1+ i \varepsilon)$ when taking $\eta_1 \rightarrow - \infty$, 
such that the time-integration is projected from $\vert 0 \rangle$ to $\vert\Omega\rangle$.
Equation \eqref{eq:free_zeta evolution} is then interpreted as the two-point correlation of $\zeta_I$ evolved from $\eta_1$ to $\eta$ after projecting the 
free vacuum to the interacting vacuum with the operator 
\begin{equation}
U_I(\eta, \eta_1) = \hat{T} e^{-i \int_{\eta_1}^{\eta}d\tilde{\eta}\; H_I(\tilde{\eta})},
\end{equation}
that satisfies $U_I(\eta, \eta_1) U_I^\dagger(\eta, \eta_1) = \mathbf{1}$.

	The prototype commutator formalism \cite{Weinberg:2005vy} can be recovered by 
	a rearrangement of the first line on the right hand side of \eqref{eq:free_zeta evolution}.
Here we rearrange the second line on the right hand side of \eqref{eq:free_zeta evolution} to obtain a double-commutator representation as
\begin{align}\label{eq:double-commutator}\nonumber
\langle \zeta^2(\eta)\rangle &= \left\langle\left( \sum_{M=0}^{\infty}i^M \int^\eta d\eta_M \cdots \int^{\eta_2}d\eta_1 
 [H_I(\eta_1), \cdots [H_I(\eta_{M}),\zeta_I(\eta)]]\right.\right) \\
&\times \left. \left( \sum_{N=0}^{\infty}i^N \int^\eta d\eta_N \cdots \int^{\eta_2}d\eta_1 
 [H_I(\eta_1), \cdots [H_I(\eta_{N}),\zeta_I(\eta)]] \right) \right\rangle.
\end{align}
The Taylor expansion of $U_I$ up to second order in  $H_I$ leads to \cite{Senatore:2009cf}:
\begin{align}
\langle \zeta^2 \rangle = \langle \zeta^2\rangle_{\rm tree} +
\langle \zeta^2 \rangle_{\rm CIM} + \langle \zeta^2 \rangle_{\rm CIS,1} + \langle \zeta^2 \rangle_{\rm CIS,2} 
\end{align}
where $\langle \zeta^2\rangle_{\rm tree}$ includes all corrections at tree-level, 
$\mathrm{CIM}$ denotes the one-loop contribution from the cut-in-the-middle diagrams 
and $\mathrm{CIS}$ denotes the one-loop contribution from the cut-in-the-side diagrams.
In particular, these one-loop diagrams are given by the cubic interactions $H_I^{(3)}$ and the quartic interactions $H_I^{(4)}$ as
\begin{align}
\langle \zeta^2 \rangle_{\rm CIM} &= - \int^\eta d\eta_1 \int^\eta d\tilde{\eta}_1 
 \left\langle \left[H_I^{(3)}(\eta_1),\zeta_I(\eta)\right]  \left(\left[H_I^{(3)}(\tilde{\eta}_1),\zeta_I(\eta)\right]\right)^\dagger \right\rangle, \\
\langle \zeta^2 \rangle_{\rm CIS,1} &= -2\; \mathrm{Re} \left[ \int^\eta d\eta_2 \int^{\eta_2} d\eta_1 
\left\langle \left[  H_I^{(3)}(\eta_1), \left[H_I^{(3)}(\eta_2),\zeta_I(\eta)\right] \right] \zeta_I(\eta) \right\rangle \right] ,
\\ 
\langle \zeta^2 \rangle_{\rm CIS,2} &= -2\; \mathrm{Im} \left[  \int^\eta d\eta_1
\left\langle \left[H_I^{(4)}(\eta_1), \zeta_I(\eta)\right]\zeta_I(\eta) \right\rangle \right].
\end{align}

\begin{figure}
	\label{Fig_2}
	\begin{center}
	\subfloat[CIM\label{subfig-CIM}]{%
				\includegraphics[width=4.3cm]{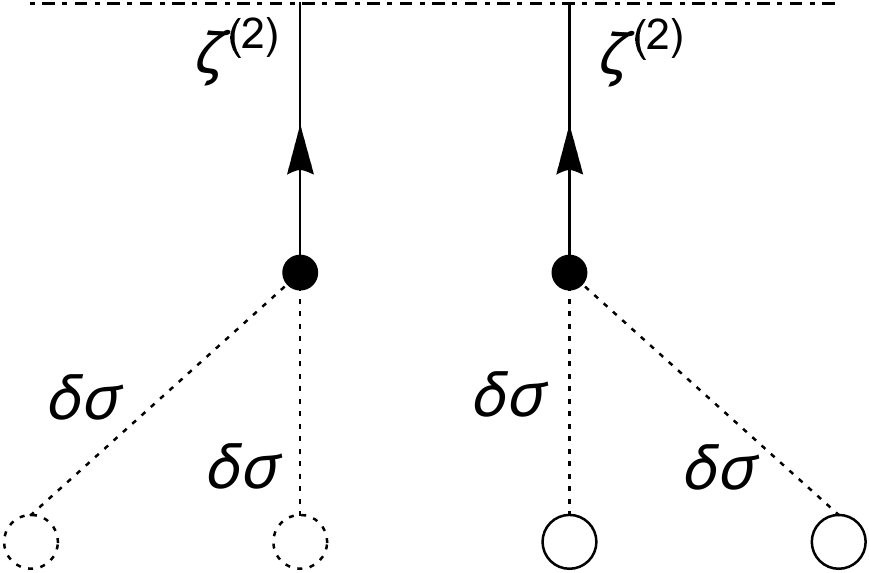}
		}	
		\hfill
		\subfloat[CIS,1\label{subfig-CIS1}]{%
			\includegraphics[width=4.3cm]{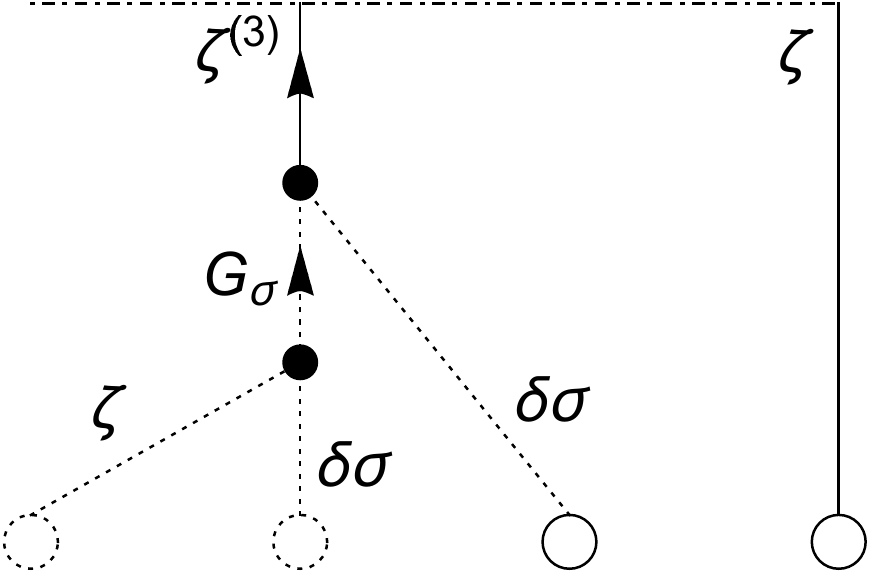}
		}
		\hfill
			\subfloat[CIS,2\label{subfig-CIS2}]{%
				\includegraphics[width=4.3cm]{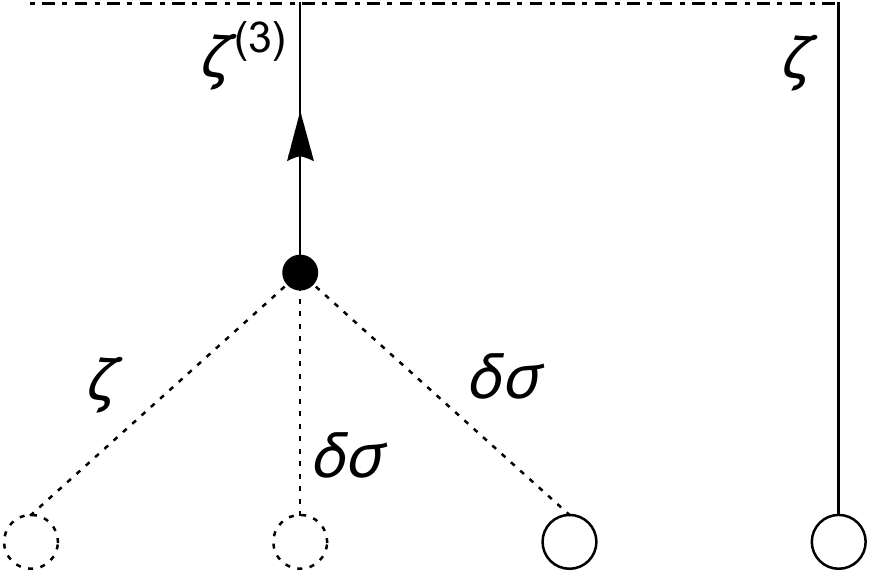}
		}
	\end{center}
	\caption{One-loop diagrams of the cut-in-the-middle ($\rm CIM$) and the cut-in-the-side ($\rm CIS$) types.
	Solid (dotted) lines are external (internal), $\zeta$ and $\delta\sigma$ are free fields, arrows are propagators, black dots are vertices,
and $\zeta^{(i)}$ is a propagating $\zeta$ field at $i$-th order in perturbations. In each diagram one dotted circle must correlates with one solid
circle in order to get non-vanished contributions, and there is no correlation between $\zeta$ and $\delta\sigma$ fields.}
\end{figure}
As seen by \eqref{sol:flat slicing}, interactions with each one $\alpha$ is suppressed by one factor $\epsilon_\phi\simeq\epsilon_H$,
and interactions involved with $\beta$ are higher-order in the slow-roll parameters. 
To the leading order in $\epsilon_\phi$, 
$H_I^{(3)}$ derived from \eqref{eq:cubic terms} comprises a $\zeta$ field and a pair of $\sigma$ fields,
while $H_I^{(4)}$ includes  a pair of $\zeta$ fields and a pair of $\sigma$ fields.
In this case there is only one corresponding $\mathrm{CIM}$ diagram, where two free $\sigma$ fields scatter to generate a propagating
$\zeta$ at second order and $\langle \zeta^2 \rangle_{\rm CIM}$ is the correlation of two of these propagating $\zeta$ fields.

There are two kinds of $\mathrm{CIS}$ diagrams. The first one is a propagating second-order $\sigma$ field generated by a scattering
of a free $\zeta$ with a free $\sigma$. This propagating $\sigma$ scatters with a free $\sigma$ field again and results in a propagating 
$\zeta$ at third order. $\langle \zeta^2 \rangle_{\rm CIS,1}$ is then the correlation of a third-order $\zeta$ with a free $\zeta$ field.
Similarly, $\langle \zeta^2 \rangle_{\rm CIS,2}$ is also a correlation of a third-order $\zeta$ with a free $\zeta$ field, where the
third-order $\zeta$ field is generated by the scattering of a free $\zeta$ with a pair of free $\sigma$ fields through the interaction $H_I^{(4)}$.

\subsection{Linear perturbations  with phase transitions} \label{Sec 4.1}
We have seen that $\sigma$ can obtain a time-dependent mass square $M^2 = V_{\sigma\sigma}/H^2$ during a transition 
to another local minimum, and that $M^2$ can become negative during phase transitions.
To specify the discussion, we consider a transition driven by the type of potentials as \eqref{eq:potential_sigma}.
We can solve the mode function $u_k = a \delta\sigma_k$ with a simplified equation of motion given by
\begin{equation}\label{eq:eom_mode}
u_k^{\prime\prime}+\left( k^2 - \frac{l^2 -1/4}{\eta^2}\right) u_k =0, \;\;\;\;
l = \left\lbrace 
\begin{array}{ll}
3/2, \;\;&\; \eta\leq \eta_i, \\
\sqrt{9/4 + M_-^2}, \;\;&\; \eta_i < \eta \leq \eta_f, \\
\sqrt{9/4 - M_+^2}, \;\;&\; \eta > \eta_f,
\end{array} \right.
\end{equation}
where $M_-^2=\vert M^2\vert$ is the phase that $M^2 < 0$, and therefore $l > 3/2$ for $\eta_i < \eta \leq \eta_f$ and
$0 < l < 3/2$ for $\eta > \eta_f$.
Without the loss of generality, $\sigma$ is assumed to be massless by $\eta_i$.

Given that $\sigma$ starts to roll down from very close to the origin,
we may neglect the change of $\sigma$ in \eqref{def:M^2} 
so that $M^2\approx - \lambda v^2/H^2$ is a good approximation in the first few e-foldings,
where the time dependence in $M$ is mild with a sufficently small $\epsilon_H$. 
To reach a good understanding on the behavior of $u_k$, it is enough to consider a pure de Sitter expansion 
by taking $\epsilon_H=0$, and thus $H = H_i$, $\eta = - 1/\mathcal{H} = - 1/(H_i a)$. 
Using a new time variable $z = k \eta$, 
the solution of \eqref{eq:eom_mode} with a constant $M_-^2=  \lambda v^2/H_i^2$ and a constant $M_+^2 = 2\lambda v^2/H_i^2$ reads
\begin{align} \label{sol:sigma_k phase1}
\delta\sigma_k &= \frac{H}{\sqrt{2 k^3}}e^{-i z}(i- z), &z \leq z_i,\\ \label{sol:sigma_k phase2}
&= \frac{H}{\sqrt{ k^3}}(- z)^{3/2}[b_1(k) J_{l_-}(-z) + i\, b_2(k) Y_{l_-}(-z)], &  z_i < z \leq z_f, 
\\ \label{sol:sigma_k phase3}
&= \frac{H}{\sqrt{ k^3}}(- z)^{3/2}[c_1(k) J_{l_+}(-z) + i\, c_2(k) Y_{l_+}(-z)], &  z > z_f,
\end{align}
where $z_i\equiv k\eta_i$, $l_\pm \equiv \sqrt{9/4 \mp M^2_\pm}$, and $J_l (Y_l)$ is the Bessel fuction of the first (second) kind.

Let us first focus on the time interval $z_i < z \leq z_f$.
If $ M_-^2 \ll 1$, the mode function $\delta\sigma_k$ with $k< k_i\equiv -1/\eta_i$ is nearly a constant by $z=z_i$.
In the limit $z\rightarrow 0$, $\delta\sigma_k\rightarrow iH/\sqrt{2k^3}$ in \eqref{sol:sigma_k phase1} while 
$J_l(-z)$ automatically drops out in \eqref{sol:sigma_k phase2}, and one can match the solution at $z = z_i$ to find that
$b_2 = -\pi (-z_i)^{l-3/2}/(\sqrt{2}\,2^l\Gamma(l))$, where
\begin{equation}
\label{sol:sigma_k small mass}
\delta\sigma_k = i \frac{H}{\sqrt{2k^3}}\left(\frac{z}{z_i}\right)^{3/2-l}.
\end{equation}
On the other hand, $\delta\sigma_k$ with $k> k_i$ leaves the horizon at some epoch $\eta_k > \eta_i$, where $\eta_k \gtrsim -1/k$.
Although $b_1$, $b_2$ can be solved by matching the value of $\delta\sigma_k$ and $\delta\sigma_k^\prime$ at $z= z_i$, 
in the limit $z_i\gg 1$ we find that
\begin{equation}
b_1 \approx b_2 \rightarrow \frac{\sqrt{\pi}}{2}\exp\left[i\left(\frac{\pi}{2}l + \frac{\pi}{4}\right)\right],
\end{equation}
which are independent of $z_i$. This implies that mode functions with $k \gg k_i$ cannot feel the phase transition,
and the vacuum state in the Minkowski regime $\delta\sigma_k\approx - (H/\sqrt{2k^3})\times z e^{-i z}$ is almost unchanged after $z_i$.
Therefore, for $k> k_i$ we shall match $\vert\delta\sigma_k\vert$ between the vacuum state with 
 the late-time growing mode $Y_l$ in \eqref{sol:sigma_k phase2}
at $z = z_k\equiv k\eta_k$, which leads to
\begin{equation}
\label{sol:sigma_k k>k_i}
\delta\sigma_k = i \frac{H}{\sqrt{2k^3}}z_k\left(\frac{z}{z_k}\right)^{3/2-l}.
\end{equation}

For $\vert M^2\vert > 1$, the time evolution of $\delta\sigma_k$ becomes important after the phase transition.
As a simple example, let us consider a special case $l = 5/2$ (that is $M_-^2 = 4$ ), where \eqref{sol:sigma_k phase2} has an
analytical expression based on $J_{5/2}(-z)$ and $Y_{5/2}(-z)$. 
By matching $\delta\sigma_k$ and $\delta\sigma_k^\prime$ at $z= z_i$, we can solve the coefficients as 
\begin{align}
b_1 &= \frac{\sqrt{\pi}}{4 z_i^4}\left[e^{-2i z_i}\left(3i -6z_i -2i z_i^2\right) + 3i +2z_i^2\left(2i+ z_i(2-i z_i)\right)\right],\\
b_2 &= \frac{\sqrt{\pi}}{4 z_i^4}\left[e^{-2i z_i}\left(-3i +6z_i +2i z_i^2\right) + 3i +2z_i^2\left(2i+ z_i(2-i z_i)\right)\right],
\end{align}
and now \eqref{sol:sigma_k phase2} is given by
\begin{align}
\nonumber
\delta\sigma_k = -& \frac{H}{2\sqrt{2 k^3}z_i^4 z}e^{-i z}\left[(z(z-3i)-3)(2z_i^2(z_i(z_i+2i)-2)-3)\right. \\
\label{sol:sigma_k 5/2}
&- \left. e^{2i(z-z_i)}(z(z+3i)-3)(2z_i(z_i-3i)-3) \right].
\end{align}
Considering the limit $\vert z\vert \ll \vert z_i\vert \leq 1 $ for mode functions with $k \leq k_i$, 
\eqref{sol:sigma_k 5/2} is reduced as
\begin{equation}
\delta\sigma_k \approx\frac{4 i}{5}\frac{H}{\sqrt{2k^3}}\left(\frac{z_i}{z}\right).
\end{equation}
This solution agrees with \eqref{sol:sigma_k small mass} for $l=5/2$ up to a constant factor $4/5$.
\footnote{In the limit $ z_i \leq z \ll -1 $, 
	\eqref{sol:sigma_k 5/2} reproduces the usual Bunch-Davies vacuum state
	$\delta\sigma_k\approx - (H/\sqrt{2k^3})\times z e^{-i z}$ in the Minkowski regime. 
}

Similarly, one can solve $c_1$, $c_2$ in \eqref{sol:sigma_k phase3} by matching the solutions at $z = z_f$.
For $k < k_i$, we can simply match \eqref{sol:sigma_k small mass} with \eqref{sol:sigma_k phase3} in the limit $z\rightarrow 0$ 
and find that
\begin{equation}\label{eq:sigma_k phase3 latetime}
\delta\sigma_k \rightarrow i \frac{H}{\sqrt{2k^3}}\left(\frac{z_f}{z_i}\right)^{l_+- l_-}\left(\frac{z}{z_i}\right)^{3/2-l_+}.
\end{equation}
Since $l_+ < 3/2$, the mode function decays for $z > z_f$ as the case of a massive field with a positive mass square \cite{Weinberg:2006ac}.
The general solution of the mode function $\delta\sigma_k$ with phase transitions at $z_i$ and $z_f$ can be found in appendix \ref{Appendix C}. 
 
\subsection{Induced curvature perturbations} 

\subsubsection{tree-level contributions}
\begin{figure}
	\begin{center}
		\includegraphics[width=3.7cm]{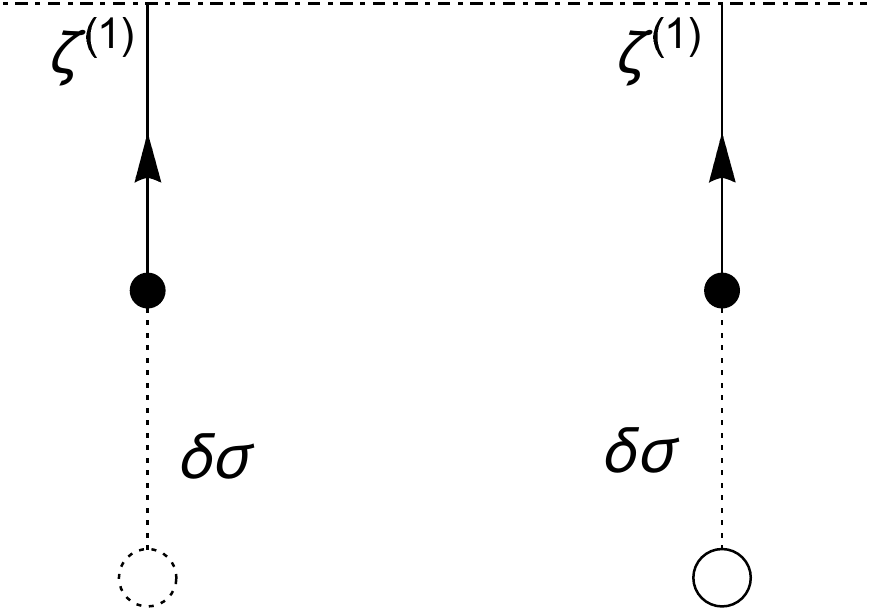}
		\qquad\qquad\qquad
		\includegraphics[width=3.7cm]{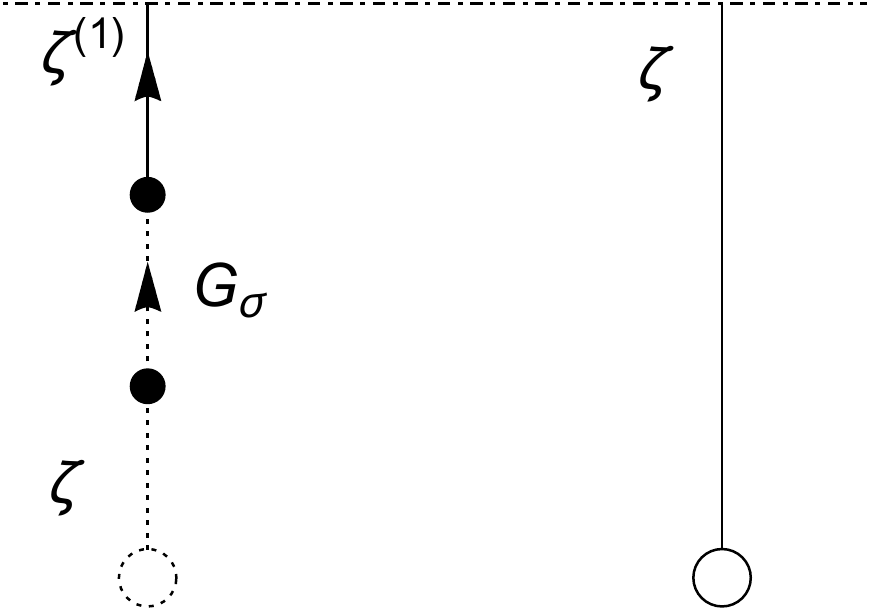}
	\end{center}
	\caption{	\label{Fig_tree} Leading tree-level diagrams of the cut-in-the-middle ($\rm CIM$) and the cut-in-the-side ($\rm CIS$) types.
		Solid (dotted) lines are external (internal), $\zeta$ and $\delta\sigma$ are free fields, arrows are propagators, black dots are vertices,
		and $\zeta^{(i)}$ is a propagating $\zeta$ field at $i$-th order in perturbations. In each diagram one dotted circle must correlates with one solid
		circle in order to get non-vanished contributions, and there is no correlation between $\zeta$ and $\delta\sigma$ fields.}
\end{figure}

We can easily learn the advantage of using the double-commutator formalism \eqref{eq:double-commutator} by starting the calculation at tree-level.
Since in the spatially flat slicing $\delta\sqrt{-g} = a^4 \alpha \approx - \epsilon_\phi a^4 \zeta$ at linear order, all of the quadratic interactions
$H_I^{(2)}$ coming from the gravitational couplings must exist as a pair. 
Up to second order in $H_I^{(2)}$, there are two kinds of corrections to the curvature perturbation given by 
\begin{align}\label{eq:tree-corrections}\nonumber
	\langle \zeta^2(\eta)\rangle_{\rm tree} = \langle 0\vert \zeta_I^2(\eta) \vert 0\rangle 
	+\langle \zeta^2(\eta)\rangle^{\rm tree}_{\rm CIM} + \langle \zeta^2(\eta)\rangle^{\rm tree}_{\rm CIS},
\end{align}
where $\langle 0\vert \zeta_I^2 \vert 0\rangle$ is the tree-level contribution from inflaton, and we define
\begin{align}
\langle \zeta^2\rangle^{\rm tree}_{\rm CIM} =&-\int^\eta d\eta_1 \int^\eta d\tilde{\eta}_1 
\left\langle \left[H_I^{(2)}(\eta_1),\zeta_I(\eta)\right]  \left(\left[H_I^{(2)}(\tilde{\eta}_1),\zeta_I(\eta)\right]\right)^\dagger \right\rangle, \\
\langle \zeta^2\rangle^{\rm tree}_{\rm CIS} =& -2\; \mathrm{Re} \left[ \int^\eta d\eta_2 \int^{\eta_2} d\eta_1 
\left\langle \left[  H_I^{(2)}(\eta_1), \left[H_I^{(2)}(\eta_2),\zeta_I(\eta)\right] \right] \zeta_I(\eta) \right\rangle \right] .
\end{align}
Here $\langle \zeta^2\rangle^{\rm tree}_{\rm CIM}$ is the correlation of two propagating $\zeta$ fields converted from two free $\sigma$ fields.
$\langle \zeta^2\rangle^{\rm tree}_{\rm CIS}$ is the correlation of one free $\zeta$ field with one propagating $\zeta$ field, 
where the propagating $\zeta$ was a free $\zeta$ in the first place but it converted once to a propagating $\sigma$ at $\eta_1$
and converted back at $\eta_2$ (see Fig. \ref{Fig_tree}).

In our scenario, the non-derivative interactions due to $V_1(\sigma)$ has time dependence.
We impose the same assumption as for solving the mode function \eqref{eq:eom_mode}, and the quadratic interaction
of our interest thus has three phases, which are
\begin{align}\label{eq:quadratic_interaction phase1}
	\mathcal{H}_I^{(2)}= -\mathcal{L}_I^{(2)} &= \epsilon_\phi \frac{a^4}{2} g^3 \zeta \delta\sigma,  &\eta\leq \eta_i, 
	\\\label{eq:quadratic_interaction phase2}
	&= \epsilon_\phi \frac{a^4}{2} \lambda v^2 \sigma \zeta \delta\sigma,  &\eta_i < \eta \leq \eta_f, 
	\\\label{eq:quadratic_interaction phase3}
	&= \epsilon_\phi \frac{a^4}{2} \lambda(v^2 \sigma -\sigma^3) \zeta\delta\sigma, & \eta > \eta_f.
\end{align}
In the double-commutator formalism, the propagators always evolve one direction in time.
For convenience, we split every time-integration into three parts: 
\begin{align}\label{def:t-integration split}
	\mathcal{I} = \int_{-\infty}^{\eta_0} =
	\int_{-\infty}^{\eta_i}+ \int_{\eta_i}^{\eta_f} +\int_{\eta_f}^{\eta_0} 
	= \mathcal{I}_1 +\mathcal{I}_2 +\mathcal{I}_3.
\end{align}

If there is no phase transitions (that is $\eta_i\rightarrow\eta_f\rightarrow \eta_0$), then $\mathcal{I}_1$ shows the tree-corrections of
a massless field with a linear potential. According to \eqref{eq:quadratic_interaction phase1}, $\mathcal{I}_1$ is suppressed as
$g\rightarrow 0$. On the other hand, $\mathcal{I}_3$ accounts for the tree-corrections of a massive field with a positive mass square,
and therefore $\mathcal{I}_3 \rightarrow 0$ as $\sigma \rightarrow v$. Both contributions from $\mathcal{I}_1$ and $\mathcal{I}_3$
are less important since $\delta\sigma$ is either a constant or decaying function during the time.

We shall focus on contributions from $\mathcal{I}_2$ as not only $\delta\sigma$ but also $\sigma$ are growing with time during this phase
\footnote{The growing of the homogeneous value $\sigma(\eta)$ for $\eta_i < \eta < \eta_f$ 
	is nevertheless due to the choice of the origin in the potential \eqref{eq:potential_sigma}.}.
As an example, we compute
\begin{align}  
\label{eq:tree_zeta_CIM}
\langle \zeta_\mathbf{k}(\eta)\zeta_\mathbf{K}(\eta)\rangle^{\rm tree}_{\rm CIM} \supset & \left(\frac{\lambda v^2}{2}\right)^2
\int_{\eta_i}^{\eta_f} d\eta_1 \int_{\eta_i}^{\eta_f} d\tilde{\eta}_1 
(2\epsilon_1) a^4(\eta_1) \sigma(\eta_1) (2\tilde{\epsilon}_1) a^4(\tilde{\eta}_1) \sigma(\tilde{\eta}_1)\nonumber\\
&\times G^\zeta_{\mathbf{k}}(\eta ; \eta_1) G^\zeta_\mathbf{K}(\eta ; \tilde{\eta}_1)
\langle \delta\sigma_\mathbf{k}(\eta_1)\delta\sigma_\mathbf{K}(\tilde{\eta}_1)\rangle.
\end{align}
The $\zeta$ Green function is obtained according to \eqref{eq:mode_zeta} as
\begin{align}\label{eq:Green_zeta}
G^\zeta_{\mathbf{k}}(\eta ; \tilde{\eta}) & =  i \theta(\eta - \tilde{\eta}) 
\frac{\kappa^2 H^2}{2\sqrt{\epsilon_\phi \tilde{\epsilon}_\phi}} \frac{1}{2k^3} \\
&\qquad \times\left[e^{-ik(\eta-\tilde{\eta})}(i-k\eta)(-i-k\tilde{\eta})-e^{-ik(\tilde{\eta}-\eta)}(i-k\tilde{\eta})(-i-k\eta)\right],
\nonumber
\end{align}
where $\tilde{\epsilon}_\phi$ is evaluated at $\tilde{\eta}$. 

For $\vert k \eta\vert < 1$, we adopt the mode function \eqref{sol:sigma_k small mass} in the late-time limit with the new time variable $z \equiv k\eta$. 
Taking $\sigma \rightarrow \sigma_+ (a/a_i)^{l_- -3/2}$ from the case with $g > 0$, we find
\begin{align}  
\label{eq:tree_zeta_CIM2}
\langle \zeta_\mathbf{k}(\eta)\zeta_\mathbf{K}(\eta)\rangle^{\rm tree}_{\rm CIM} \supset & \frac{\kappa^4 H^4}{8 k^3}M_-^4
\delta(\mathbf{k} +\mathbf{K}) \left(\frac{\sigma_+}{H}\right)^2
\lvert F_4(z_f) - F_4(z_i)\rvert^2.
\end{align}
Here, we have defined a useful function
\begin{align}\label{def:F_n}
F_n(z)\equiv \int \frac{d\tilde{z}}{2 \tilde{z}^n}\Theta(z-\tilde{z}) 
\left[ e^{i(z-\tilde{z})}(1-i z)(\tilde{z}-i)+ e^{i(\tilde{z}-z)}(1+ i z)(i + \tilde{z})\right]
\left(\frac{\tilde{z}}{z_i}\right)^{3-2l}.
\end{align}
In the limit $\vert z \vert \leq \vert \tilde{z}\vert \ll 1$, the time integration is simplified as
\begin{align}
\label{eq:simple_F}
F_n(z)\approx \int \frac{d\tilde{z}}{3 \tilde{z}^n}\Theta(z-\tilde{z}) 
\left( z^3 - \tilde{z}^3\right)
\left(\frac{\tilde{z}}{z_i}\right)^{3-2l}.
\end{align}
Thus in the late-time limit \eqref{eq:tree_zeta_CIM2} reads
\begin{align}  
\label{eq:tree_zeta_CIM3}
\langle \zeta_\mathbf{k}(\eta)\zeta_\mathbf{K}(\eta)\rangle^{\rm tree}_{\rm CIM} \supset & \frac{\kappa^4 H^4}{8 k^3}M_-^4
\delta(\mathbf{k} +\mathbf{K}) \left(\frac{g^3}{6l_- H^3}\right)^2 \Pi_\zeta(z_f),
\end{align}
where the integral factor with $z \rightarrow 0$ is approximately given by
\begin{align}
\label{eq:Pi}
\Pi_\zeta(z) &\approx \frac{1}{4l_-^2(3-2l_-)^2}\left(\frac{z}{z_i}\right)^{6-4l_-}, &\mathrm{if}\;\;\; \frac{3}{2}< l_- \leq \frac{5}{2}, \\
&\approx \frac{1}{9} \ln^2 \left(\frac{z_i}{z}\right), &\mathrm{if}\;\;\; l_- = \frac{3}{2}. \qquad
\end{align}
We will comment on this finding together with the results from one-loop calculations.
One can easily compute the $\mathcal{I}_2$ contribution in $\langle \zeta^2\rangle^{\rm tree}_{\rm CIS}$ by using the same method.
However, in the ${\rm CIS}$ case the two free $\delta\sigma$'s in \eqref{eq:tree_zeta_CIM} are replaced by $\zeta$'s
(and with one $G_\zeta$ being replaced by one $G_\sigma$),
and therefore only the evolution of $\sigma$ gives temporal enhancement in the late-time limit.
This fact implies that the $\mathcal{I}_2$ contribution in $\langle \zeta^2\rangle^{\rm tree}_{\rm CIS}$ 
$\sim (z_f/z_i)^{3-2l_-}$, which is clearly a subdominant contribution than $\Pi_\zeta(z_f)\sim (z_f/z_i)^{6 - 4l_-}$.

\subsubsection{one-loop contributions}
We now calculate the one-loop corrections to the $\zeta$ two-point function induced by the $\sigma$ field 
which undergoes a phase transition as \eqref{eq:eom_mode}. 
The late-time contribution from $\langle \zeta^2 \rangle_{\rm CIM}$ involves two propagators of each
$G_\zeta\sim a^{-3}$, four free fields of each $\delta\sigma\sim a^{l - 3/2}$ and two vertices. 
For the non-derivative cubic interaction \eqref{def:nonderivative_interaction}
each vertex brings in an integration $\sim \epsilon_\phi  V_{\sigma\sigma}\int d\eta a^4(\eta) $, and we can find from \eqref{eq:zeta1_latetime} that
the $\mathcal{I}_2$ contribution in $\langle \zeta^2 \rangle_{\rm CIM} \sim a^{4l - 6}$, 
which is growing during $\eta_i < \eta \leq \eta_f$ where $l = l_- > 3/2$.
On the other hand, in the $\langle \zeta^2 \rangle_{\rm CIS,1}$ correlation, two temporarily growing $\delta\sigma$ fields 
are replaced by one propagator $G_\sigma$. 
The internal propagator $G_\sigma (\eta_1; \eta_2)$ effectively contributes $a^{-3}$ to the two-point function after both $\eta_1$ and $\eta_2$
are integrated out.
 In $\langle \zeta^2 \rangle_{\rm CIS,2}$ the quartic interactions can provide at best two free $\delta\sigma$'s with
one non-derivative vertex $\sim\epsilon_\phi^2 V_{\sigma\sigma}\int d\eta a^4(\eta)$.
In both $\rm CIS$ diagrams one of the external field is a free $\zeta$ whose contribution can be locally absorbed into the homogeneous background
through a rescaling of the scale factor $a$ \cite{Senatore:2009cf}
\footnote{ This is due to the fact that we are only interested in the case where all corrections from $\sigma$ are perturbatively small and are computed by interactions $H_I$, the renormalization
	of the scale factor $a \rightarrow a e^{\zeta_I}$ becomes independent of $\sigma$.}. 
This is the essential difference between $\rm CIS$ diagrams and the $\rm CIM$ one.
As a result one concludes that the $\mathrm{CIM}$ diagram is the dominant one-loop corrections on superhorizon scales.

We divide the time-integration in the same ways as \eqref{def:t-integration split}.
There is no loop contribution from $\mathcal{I}_1$ since $\sigma$ is massless for $\eta \leq \eta_i$.
The one-loop contributions from $\mathcal{I}_3$ are based on a decaying mode function \eqref{eq:sigma_k phase3 latetime} in the late-time limit
(see also appendix \ref{Appendix C}). For modes that have crossed the horizon by $\eta_f$, their contributions are suppressed in $\mathcal{I}_3$.
For modes that exit the horizon after $\eta_f$, their contributions in $\mathcal{I}_3$ are fixed around the time of horizon-crossing and thus
are irrelevant to the phase transition.

Let us compute the $\mathcal{I}_2$ contribution in $\langle \zeta^2 \rangle_{\rm CIM}$.
The leading interactions of $\mathcal{O}(\epsilon_\phi)$ from \eqref{eq:cubic terms} are
\begin{equation}\label{eq:leading interactions}
S^{(3)}_{\zeta\sigma\sigma} \supset \int d\eta d^{3}x \frac{a^{2}}{2}\alpha\left[- \delta\sigma^{\prime 2} -(\partial\delta\sigma)^2 -a^2 V_{\sigma\sigma}\delta\sigma^2\right],
\end{equation} 
where after integrating the first term with time derivatives by parts and replacing it by the equation of motion \eqref{eq:eom_free sigma_k}, 
we can recast the interactions \eqref{eq:leading interactions} as  
\begin{equation}\label{eq:leading interactions2}
S^{(3)}_{\zeta\sigma\sigma} \supset \int d\eta d^{3}x \frac{a^{2}}{2}\alpha\left[\delta\sigma \partial^2\delta\sigma -(\partial\delta\sigma)^2 -2 a^2 V_{\sigma\sigma}\delta\sigma^2 + \mathcal{O}(\epsilon_\phi) \dots\right].
\end{equation}
By virtue of the Green's method for cosmological correlators \cite{Musso:2006pt}, we treat the $\delta\sigma$ fields as
additional sources to the $\zeta$ equation of motion as
\begin{equation}\label{eq:sourced eom zeta}
\zeta_\mathbf{k}^{\,\prime\prime} + (2 + \frac{\epsilon_\phi^\prime}{\mathcal{H}\epsilon_\phi}) \mathcal{H}\zeta_{\mathbf{k}}^{\, \prime} 
+k^2\zeta_\mathbf{k}^{} = \mathcal{S}_\zeta(\mathbf{k},\eta),
\end{equation}
where we define
\begin{align}\label{eq:source total}
S_\zeta(\mathbf{k},\eta) = \frac{\kappa^2}{4} \int\frac{d^3\mathbf{p}d^3\mathbf{q}}{(2\pi)^3}\delta(\mathbf{p}+\mathbf{q}-\mathbf{k}) \delta\sigma_\mathbf{p}(\eta)\delta\sigma_\mathbf{q}(\eta) 
 \left[-2\frac{M^2}{\eta^2}+q^2-p_iq^i +\dots  \right].
\end{align}
The correlation function due to the $\mathrm{CIM}$ diagram is then given by
\begin{align} \label{eq:co_zeta total}
\langle \zeta_\mathbf{k}(\eta)\zeta_\mathbf{K}(\eta)\rangle_{\rm CIM} =  &
\int d\tilde{\eta}_2 \int d\tilde{\eta}_1 (2\tilde{\epsilon}_1) a^2(\tilde{\eta}_1) (2\tilde{\epsilon}_2) a^2(\tilde{\eta}_2) \nonumber\\
&\times G^\zeta_{\mathbf{k}}(\eta ; \tilde{\eta}_1) G^\zeta_\mathbf{K}(\eta ; \tilde{\eta}_2)
\langle S_\zeta(\mathbf{k},\tilde{\eta}_1)S_\zeta (\mathbf{K},\tilde{\eta}_2)\rangle,
\end{align}
where $\epsilon_1\equiv\epsilon(\eta_1)$ and 
$G^\zeta_{\mathbf{k}}$ is the Fourier transform of the Green function $G_\zeta$. The unequal-time correlation function is
\begin{align}\label{eq:source pair}
\langle & S_\zeta(\mathbf{k},\eta_1) S_\zeta (\mathbf{K},\eta_2)\rangle = \frac{\kappa^4}{16}
\int\frac{d^3\mathbf{p}d^3\mathbf{q} d^3\mathbf{P} d^3\mathbf{Q}}{(2\pi)^6} 
\delta(\mathbf{p}+\mathbf{q}-\mathbf{k})\delta(\mathbf{P}+\mathbf{Q}-\mathbf{K}) \nonumber\\
&\times \left[  -2\frac{M^2}{\eta_1^2} +\dots \right]\left[  -2\frac{M^2}{\eta_2^2}+\dots \right]
\langle \delta\sigma_\mathbf{p}(\eta_1)\delta\sigma_\mathbf{q}(\eta_1) \delta\sigma_\mathbf{P}(\eta_2)\delta\sigma_\mathbf{Q}(\eta_2) \rangle,
\end{align}
where $M = M_-$ for $\eta_i < \eta \leq \eta_f$ and $M = M_+$ for $\eta > \eta_f$.
To simplify the calculation, we assume that the slow-roll parameter is nearly
a constant throughout the time of our interest.
By using the symmetry between the pairs of momentums
$(\mathbf{p},\mathbf{q})\leftrightarrow (\mathbf{P},\mathbf{Q})$ and the symmetry 
between time integrations ($\tilde{\eta}_1 \leftrightarrow \tilde{\eta}_2$), we can rearrange the integration as
\begin{align}  
\label{eq:co_zeta_CIM}
\langle \zeta_\mathbf{k}(\eta)\zeta_\mathbf{K}(\eta)\rangle_{\rm CIM} = &
\frac{\kappa^4 M^4}{2} \delta(\mathbf{k} +\mathbf{K}) \int d^3\mathbf{p}d^3\mathbf{q} \delta(\mathbf{p}+\mathbf{q}-\mathbf{k}) \nonumber\\
&\;\;\;\;\;\times \left| \int_{}^{\eta}\frac{d\tilde{\eta}}{\tilde{\eta}^2}a^2(\tilde{\eta})G^\zeta_\mathbf{k}(\eta ; \tilde{\eta})
\delta\sigma_\mathbf{p}(\tilde{\eta})\delta\sigma_\mathbf{q}(\tilde{\eta}) \right|^2,
\end{align}
where the pair of momentums $(\mathbf{P},\mathbf{Q})$ have been integrated out.
Note that the disconnected diagram $\langle \zeta\zeta \rangle_{dc} \equiv \langle \zeta\rangle^2$
has been excluded in \eqref{eq:co_zeta_CIM}  as it only contributes to the zero-mode correlation $k=K=0$.
\footnote{The disconnected diagram corresponds to the equal-time propagation of the source pairs in \eqref{eq:source pair}.} 

It is useful to reparametrize the internal wave numbers with respect to the external wave number as $x= q/k$, $y=p/k$.
For $k < k_i$ (or $z_i < 1$), we shall compute the time from $z= z_i$ and put the cutoff $y_i\equiv k_i/k$ to the internal wave numbers so that
the late-time approximation \eqref{sol:sigma_k small mass} can be applied.
After applying these cutoffs we extract only the contribution from superhorizon modes since the start of the phase transition at $\eta =\eta_i$.
As we will see that the choice of the UV cutoff $y_i$ does not affect the results.
The $\mathcal{I}_2$ contribution in the correlation \eqref{eq:co_zeta_CIM} is then
\begin{align}  \label{eq:co_zeta_CIM zi<1}
\langle \zeta_\mathbf{k}\zeta_\mathbf{K}\rangle_{\rm CIM} \supset \frac{\pi \kappa^4H^4}{ k^3}M_-^4 \delta(\mathbf{k} +\mathbf{K})
\int_{y_0}^{y_i}\int_{\vert 1-y\vert}^{1+y} dxdy \frac{1}{x^2 y^2} 
\lvert F_4(z_f) - F_4(z_i)\rvert^2, 
\end{align} 
where $y_0 = k_0/k$, and $k_0$ is a suitable cutoff for the largest cosmological scale (see the discussion in \cite{Seery:2007wf,Xue:2011hm}). 
Taking the late-time approximation \eqref{eq:simple_F} into \eqref{eq:co_zeta_CIM zi<1}, we find that the dominant contribution comes from the
 IR cutoff $y_0$, where
\begin{equation}
\label{eq:CIM_result}
\langle \zeta_\mathbf{k}\zeta_\mathbf{K}\rangle_{\rm CIM} \supset
\frac{3\pi \kappa^4H^4}{ k^3 }M_-^4 \delta(\mathbf{k} +\mathbf{K}) \ln\left(\frac{k}{k_0}\right)\Pi_\zeta(z_f),
\end{equation}
and $\Pi(z)$ is given by \eqref{eq:Pi}.
We put a conservative upper bound $M_- \leq 2$ to justify the assumption \eqref{sol:sigma_k small mass}.
The logarithmic scale-dependence $\ln(k/k_0)$ in \eqref{eq:CIM_result} is subject to the projection effect between two superhorizon modes,
which should have no importance for cosmological observables \cite{Senatore:2012nq}.
The study of this issue is beyond the scope of the current paper.

\begin{figure}
	\begin{center}
		\includegraphics[width=7cm]{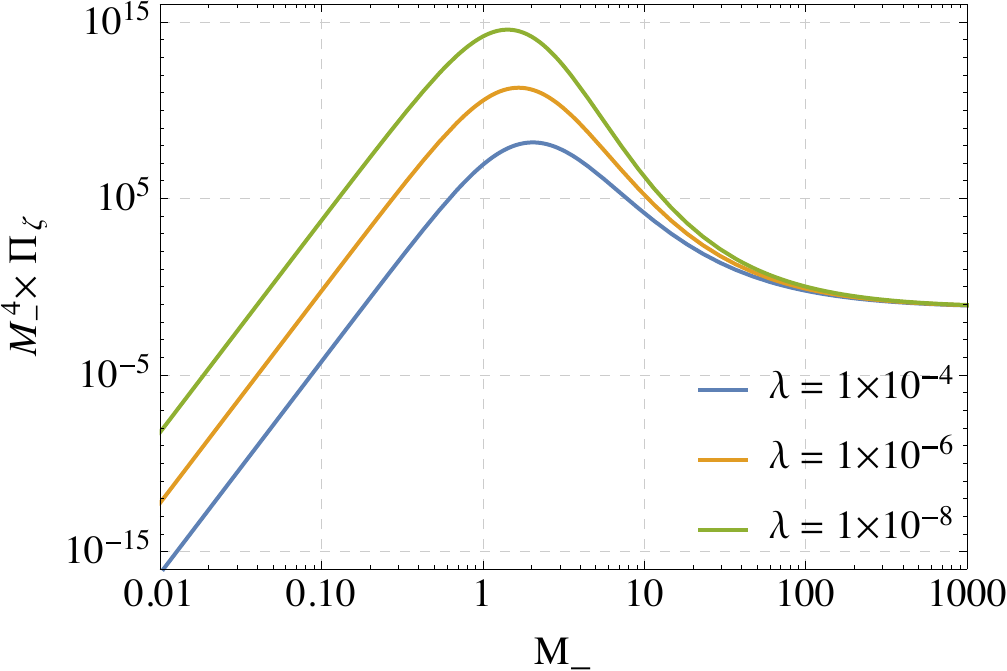}
		\includegraphics[width=6.73cm]{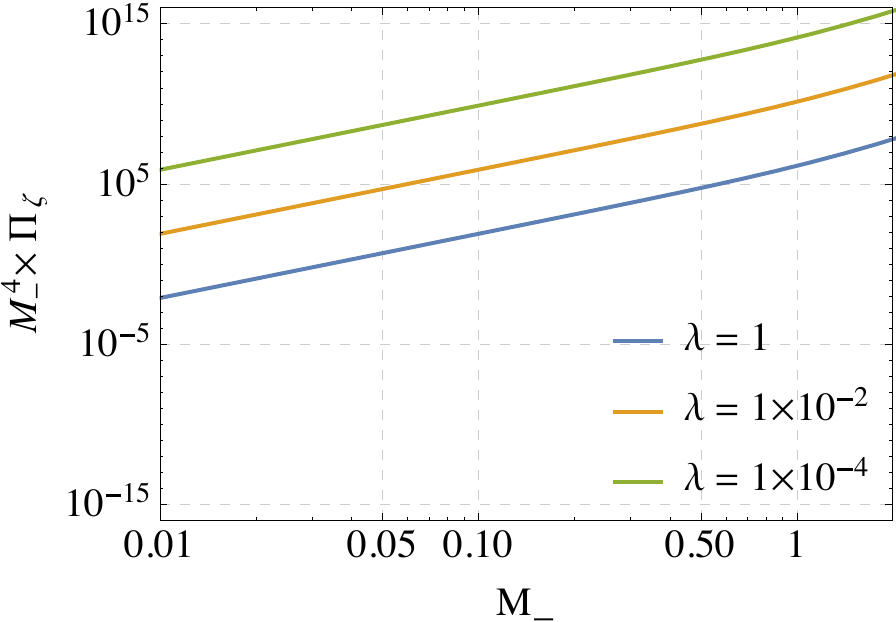}
	\end{center}
	\caption{	\label{Fig_3} The value of the mass-dependence in the one-loop $\langle \zeta^2\rangle_{\rm CIM}$ correlation function
		 for $k < k_i$ given by Eq.~\eqref{eq:CIM_result} with respect to the mass parameter $M_-$ evaluated at $\eta = \eta_f$.
		The left panel shows the case with $g=0$, and the right panel shows the case with $g = H/2$.}
\end{figure}

Let us consider the case with $g =0$ as an example.
Since $\delta\sigma_k$ starts to decay after the mass square changes to be positive, 
the source $\mathcal{S}_\zeta$ has a maximum value around $\eta = \eta_f$.
We may estimate the value of $\Pi_\zeta(z_f)$ from \eqref{eq:duration_of_slowroll} in a pure de Sitter expansion
by taking $M_-^2 = M_i^2$.
Suppose that the energy scale of inflation is $H_i^2/M_p^2\sim \epsilon_i\times 10^{-10}$, then the result in Figure~\ref{Fig_3} shows that 
the one-loop correction $\langle \zeta^2\rangle_{\rm CIM}$ can have a value around $\epsilon_i^2 M^4\Pi_\zeta \times 10^{-10}$ times smaller than the 
tree expectation value $\langle\zeta^2\rangle_{\rm tree}\sim H_i^2/(\epsilon_i M_p^2)$. 
	 With a self-coupling $\lambda\sim 10^{-8}$ and $\epsilon < 0.0068$ according to \cite{Ade:2015lrj},
one may realize $\langle \zeta^2\rangle_{\rm CIM}/\langle \zeta^2\rangle_{\rm tree} \sim \mathcal{O}(1)$,
which should be constrained by the perturbativity and unitarity conditions.
However, our purpose here is to show that the loop correction can be much larger than the usual expectation from massless fields or massive spectator fields 
that always live in some stable vacuum states \cite{Weinberg:2006ac,Senatore:2009cf}. We give a more detailed analysis on the allowed parameter space of $\lambda$ and $M_-$ in the next section together with higher-order corrections.
In Figure~\ref{Fig_3} the result of \eqref{eq:duration_of_slowroll} is used and we have checked that
this result is nearly unchanged by using the result of \eqref{eq:DeltaN_small_mass} for $M_- < 1$.
The condition $\Delta N > 0$ indicates $M_- > 0.044$ for $\lambda = 1 \times 10^{-8}$. 
Note that the decay of $\delta\sigma_k$ at $\eta > \eta_f$ only makes $\mathcal{S}_\zeta$ decay in \eqref{eq:sourced eom zeta}
but the curvature perturbation is always enhanced during the phase transitions.
As $\mathcal{S}_\zeta\rightarrow 0$, $\zeta$ converges to a constant again and the $\langle \zeta^2\rangle_{\rm CIM}$
correction enhanced by temporarily growing perturbations becomes nearly frozen from $\eta = \eta_f$ to the end of inflation.

For $k_f > k > k_i$, we have $z_i > 1$ and the time integration is chosen to start at some epoch $z_k\gtrsim -1$
where all internal modes $p\leq k$ with the cutoff $y=1$ have left the horizon.
In this case the correlation function \eqref{eq:co_zeta_CIM} with the solution \eqref{sol:sigma_k k>k_i} is
again dominated by the contribution around the IR cutoff $y_0$ as
\begin{align}  \label{eq:co_zeta_CIM zi>0}
\langle \zeta_\mathbf{k}\zeta_\mathbf{K}\rangle_{\rm CIM} &\supset \frac{\pi \kappa^4H^4}{ k^3 }M^4 \delta(\mathbf{k} +\mathbf{K})
\int_{0}^{1}\int_{ 1-y}^{1+y} dxdy \frac{1}{x^2 y^2} \nonumber\\
&\qquad\qquad\qquad\qquad\qquad\times \left(\frac{x y}{y_i^2}\right)^{3-2l}
\lvert F_4(z_f) - F_4(z_k)\rvert^2, \\
& \approx \frac{\pi \kappa^4H^4}{ k^3 }M^4 \delta(\mathbf{k} +\mathbf{K}) \frac{z_i^{6-4l}}{(l-1)(3-2l)}\Pi_\zeta(z_f),
\end{align}
for $\frac{3}{2}< l \leq \frac{5}{2}$ in the limit $z \rightarrow 0$.
The factor $z_i^{6-4l} = (k/k_i)^{6-4l}$ is introduced by the scale-dependent cutoff $z = z_k$, 
which suppressed the $\mathrm{CIM}$ contribution to modes that exit the horizon during the period $\eta_f < \eta < \eta_i$.


In the case with $g > 0$, we use the result of \eqref{eq:DeltaN g>0} to obtain the right panel of Figure \ref{Fig_3}.
In this case the enhancement due to $\Pi(z_f)$ becomes more efficient with the increase of $M_-$ so that our 
approximation breaks down for $M_- > 1$. 
	Note that although the tree-level result \eqref{eq:tree_zeta_CIM3} has a strong dependence on the parameter $g$,
the one-loop correction given by Eq.~\eqref{eq:CIM_result} is independent of $g$ \footnote{We thank the anonymous referee to point this out.}. 
The reason is that, while the linear potential $- g^3 \sigma$ in \eqref{eq:potential_sigma} determines the value $\sigma_+$ at $\eta_i$, it has no effect on the
quantum fluctuation $\delta\sigma_k$. In the case with a very small $g$, the quantum behavior of the $\sigma$ field at $\sigma < 0$ becomes important as the case with $g =0$  (see the discussion in Sec.~\ref{sec:III}), and therefore we shall find $\sigma_+$ approaches to the critical value $\sigma_i$. The condition $\sigma_+ \geq \sigma_i$ implies $g^3 \geq 9l H^3/(\pi M_-^2)$ for having a well-defined classical evolution of $\sigma$.

The results of the pure $\mathcal{I}_2$ contribution
in the one-loop ${\rm CIS}$ correlation functions are given in appendix \ref{Appendix D}.
We find that the mass-dependence of $\langle \zeta^2\rangle_{\rm CIS,1}\sim M_-^4\Pi_\zeta^{1/2}(z_f) \times \ln(z_f/z_i)$
and that the mass-dependence of $\langle \zeta^2\rangle_{\rm CIS,2}\sim M_-^2\Pi_\zeta^{1/2}(z_f) $, 
where $\Pi_\zeta^{1/2}(z_f) \sim (z_f/z_i)^{3-2l_-}$.
We have numerically checked that the mass-dependence of $\langle \zeta^2\rangle_{\rm CIM}\sim M_-^4\Pi_\zeta(z_f) $
is always the most important one-loop contribution for $M_- \leq 2$. This result is opposite to the case for spectator fields
with positive masses. The reason is that mode functions of positive masses are decaying on superhorizon scales
(as shown in section \ref{sec:II}), and thus the diagrams with more free $\delta\sigma$ fields are further suppressed in the late-time limit. 

\subsection{Second-order gravitational waves }
It is also interesting to ask how large  the corrections to the tensor perturbation $h_{ij}$ are induced by 
the (temporarily) growing field perturbations $\delta\sigma$ during phase transitions.
We are in particular interested in the contribution from the $\rm CIM$ diagram, which is the dominant effect from massive spectator fields.
The tensor perturbation induced by the $\rm CIM$ diagram is in fact a classical process that generates second-order gravitational waves
through a pair of scalar perturbations.
Unlike the curvature perturbation, the tree-level tensor mode $h$ 
does not suppress by the slow-roll parameter $\epsilon_\phi$ and therefore to find the leading contributions one can neglect any term involves with 
$\alpha$, $\beta$ or $\delta\phi$. For cubic terms the only interaction that is not suppressed by the slow-roll parameter reads \cite{Fujita:2014oba,Saito:2008jc,Saito:2009jt,Ananda:2006af,Baumann:2007zm} (see also appendix~\ref{Appendix B})
\begin{equation}\label{eq:interaction_hss}
\mathcal{L}_{h\sigma\sigma}= \frac{a^2}{2} h_{ij}\partial_i\delta\sigma\partial_j\delta\sigma.
\end{equation}
This interaction contributes to the tensor perturbation at second order according to the equation of motion
\begin{equation}
\label{eq:eom_h real space}
h_{ij}^{\prime\prime}+ 2 \mathcal{H} h_{ij}^\prime - \nabla^2 h_{ij} = -4 \perp_{ij}^{\;\;lm} S_{lm},
\end{equation}
where $\perp_{ij}^{\;\;lm}$ is the projection operator that extracts the transverse and traceless part of the source generated by \eqref{eq:interaction_hss},
that is
\begin{equation}
S_{ij}(\mathbf{x}, \eta)= \kappa^2 \partial_i\delta\sigma(\mathbf{x}, \eta) \partial_j \delta\sigma(\mathbf{x}, \eta).
\end{equation}

The Fourier transform of the tensor perturbation is defined as
\begin{equation}
h_{ij}(\mathbf{x}, \eta) = \int\frac{d^3\mathbf{k}}{(2\pi)^{3/2}}e^{i \mathbf{k}\cdot\mathbf{x}}
[h_{\mathbf{k}}^+(\eta)e^+_{ij}(\mathbf{k})+ h_{\mathbf{k}}^\times(\eta)e^\times_{ij}(\mathbf{k})],
\end{equation}
where we introduce two unit vectors $\mathbf{e}^1$, $\mathbf{e}^2$ that form orthonormal basis together with another unit vector $\mathbf{k}/k$, 
and such that
\begin{align}
e^+_{ij}(\mathbf{k}) &= \frac{1}{\sqrt{2}}[e^1_i(\mathbf{k})e^1_j(\mathbf{k})- e^2_i (\mathbf{k}) e^2_j(\mathbf{k})], \\
e^\times_{ij}(\mathbf{k}) &= \frac{1}{\sqrt{2}}[e^1_i(\mathbf{k})e^2_j(\mathbf{k}) + e^2_i(\mathbf{k})e^1_j(\mathbf{k})].
\end{align}
Therefore the equation of motion for $h_{\mathbf{k}}^+$ or $h_{\mathbf{k}}^\times$ takes the form
\begin{equation}
h_{\mathbf{k}}^{\prime\prime}+ 2 \mathcal{H}h_{\mathbf{k}}^{\prime} + k^2 h_{\mathbf{k}}= S_h(\mathbf{k}, \eta),
\end{equation}
where the source term is led by
\begin{align}\label{eq:source_h}
S_h(\mathbf{k},\eta) = -4 \kappa^2 e^{ij}(\mathbf{k}) \int\frac{d^3\mathbf{p}d^3\mathbf{q}}{(2\pi)^3}\delta(\mathbf{p}+\mathbf{q}-\mathbf{k}) \delta\sigma_\mathbf{p}(\eta)\delta\sigma_\mathbf{q}(\eta) p_iq_j  .
\end{align}

Applying the same method as used for the curvature perturbation, we can compute the tensor two-point function $\langle h_\mathbf{k}h_\mathbf{K}\rangle$
from the unequal-time correlation function
\begin{align}\label{eq:source_h pair}
\langle  S_h(\mathbf{k},\eta_1)& S_h (\mathbf{K},\eta_2)\rangle = 16\kappa^4 e^{ij}(\mathbf{k}) e^{lm}(\mathbf{K}) 
\int\frac{d^3pd^3q d^3P d^3Q}{(2\pi)^6} p_iq_j P_l Q_m  \nonumber\\
&\times \delta(\mathbf{p}+\mathbf{q}-\mathbf{k})\delta(\mathbf{P}+\mathbf{Q}-\mathbf{K})  
\langle \delta\sigma_\mathbf{p}(\eta_1)\delta\sigma_\mathbf{q}(\eta_1) \delta\sigma_\mathbf{P}(\eta_2)\delta\sigma_\mathbf{Q}(\eta_2) \rangle.
\end{align}
In terms of the reparametrized wave numbers $x= q/k$, $y=p/k$, the projection $e^{ij}p_iq_j = -e^{ij}p_ip_j $ is simply
\begin{equation}
e^{ij}(\mathbf{k}) p_ip_j= \frac{k^2 y^2}{\sqrt{2}}\left[1-\left(\frac{1+y^2 - x^2}{2y}\right)^2\right](\cos^2\varphi -\sin^2\varphi),
\end{equation}
where $\varphi$ is the azimuthal angle of $\mathbf{p}$ on the ($\mathbf{e}^1$, $\mathbf{e}^2$) plane.
The power spectrum for either polarization
($h_{\mathbf{k}}^+$ or $h_{\mathbf{k}}^\times$) is given from the definition
\begin{equation}
\langle h_\mathbf{k}(\eta)h_\mathbf{K}(\eta)\rangle = \frac{2\pi^2}{k^3} \delta(\mathbf{k} +\mathbf{K}) \mathcal{P}_h(k, \eta).
\end{equation} 

For tensor modes that exit the horizon by $\eta = \eta_i$ where $z_i = k/k_i < 1$, 
we find no IR divergence at zero mode so that the integration can run from $0$ to $y_i$ to give
\begin{align}  \label{eq:co_h zi<1}
\langle h_\mathbf{k} h_\mathbf{K}\rangle_{\rm CIM} &\supset \frac{4\pi \kappa^4H^4}{ k^3} \delta(\mathbf{k} +\mathbf{K})
\int_{0}^{y_i}\int_{\vert 1-y\vert}^{1+y} dxdy \;\frac{y^2}{x^2 }  \\
& \qquad\qquad\qquad \times \left[1-\left(\frac{1+y^2 - x^2}{2y}\right)^2\right]^2  \lvert F_2(z_f) - F_2(z_i)\rvert^2, \nonumber\\
& = \frac{64\pi \kappa^4H^4}{15 k^3} \delta(\mathbf{k} +\mathbf{K}) \left(\frac{k}{k_i}\right)^3 \Pi_h(z_f),
\end{align} 
where $F_n(z)$ has been given in \eqref{def:F_n} and the late-time approximation \eqref{eq:simple_F} is used for the second equation.
Here we only pick up the $\mathcal{I}_2$ contribution in the correlator
 \footnote{
 		Since $\delta\sigma_k$ is a massless field by $\eta_i$, we expect that the $\mathcal{I}_1$ contribution is the same as the previous findings 
\cite{Biagetti:2013kwa,Biagetti:2014asa,Fujita:2014oba} with a canonical speed of sound $c_s^2 = 1$. On the other hand, we expect that the $\mathcal{I}_3$ contribution is less important as $\delta\sigma_k$ starts to decay after $\eta_f$.
}, and the integral factor is defined as
\begin{align}
\Pi_h(z) &\approx \frac{1}{9(5-2l)^2}, &\mathrm{if}\;\;\; \frac{3}{2}\leq l < \frac{5}{2}, \\
&\approx \frac{1}{9} \ln^2 \left(\frac{z_i}{z}\right), &\mathrm{if}\;\;\; l = \frac{5}{2}. \qquad
\end{align}
We find that the second-order tensor perturbation induced by superhorizon modes ($\delta\sigma_p$ with $p < k_i$)
is dominated by the UV cutoff $y= y_i$ and is suppressed by a factor $(k/k_i)^3$.
This is not only due to the scale-dependent cutoff $y_i$ but also 
due to the fact that the leading interaction \eqref{eq:interaction_hss} is a derivative coupling with two spatial derivatives. 
We therefore conclude that massive fields cannot contribute large corrections to the tensor perturbation
even if their perturbations are growing after horizon exit due to negative masses.


\section{Higher-order corrections}
\label{sec:higher-order}

\begin{figure}
	\label{Fig_4}
	\begin{center}
		\hfill
		\subfloat[\label{subfig-CIM_q}]{%
			\includegraphics[width=4.7cm]{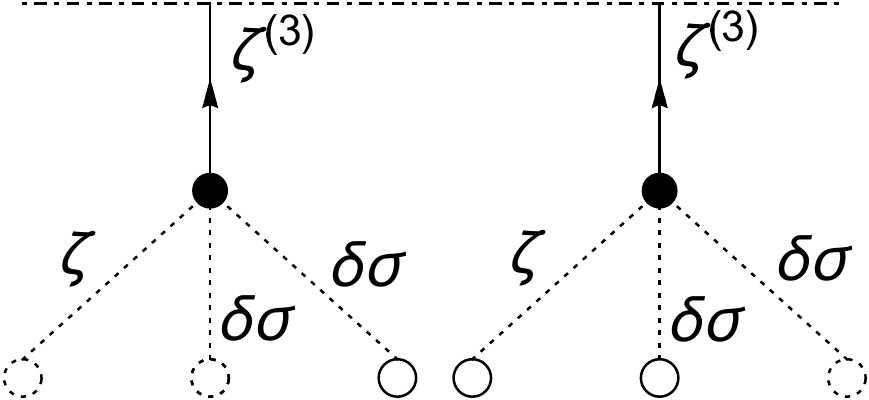}
		}
		\hfill
		\subfloat[\label{subfig-CIM_cq}]{%
			\includegraphics[width=4.7cm]{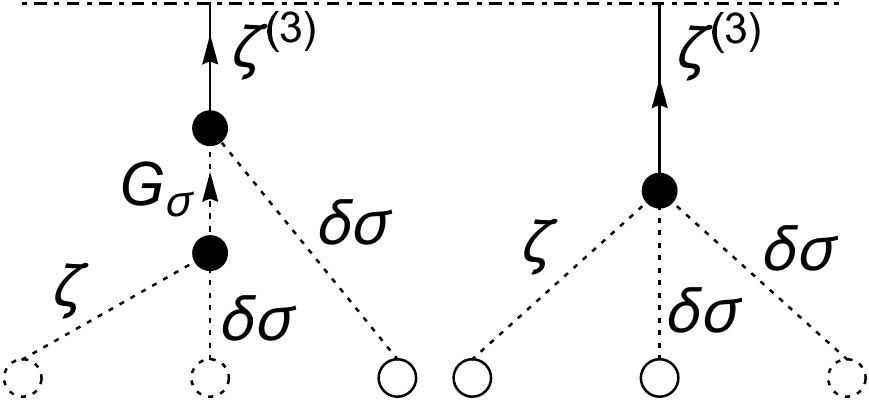}
		}
		\subfloat[\label{subfig-CIM_c}]{%
			\includegraphics[width=4.7cm]{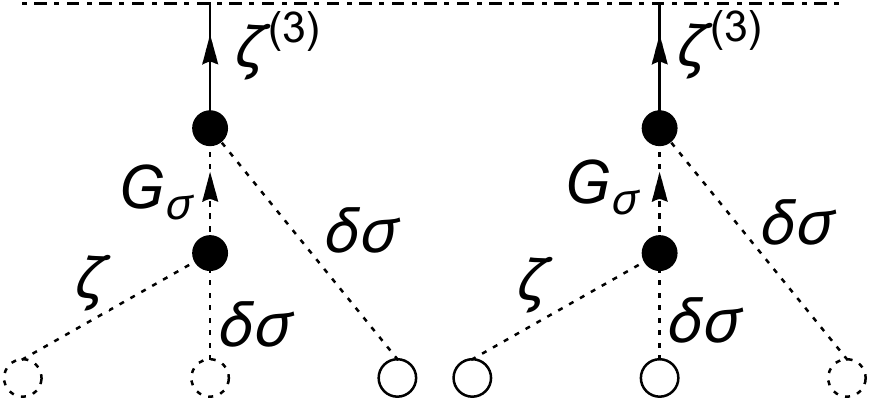}
		}		
	\end{center}
	\caption{Two-loop diagrams of the cut-in-the-middle ($\rm CIM$) type.
		Solid (dotted) lines are external (internal), $\zeta$ and $\delta\sigma$ are free fields, arrows are propagators, black dots are vertices,
		and $\zeta^{(i)}$ is a propagating $\zeta$ field at $i$-th order in perturbations. In each diagram one dotted circle must correlates with one solid
		circle in order to get non-vanished contributions, and there is no correlation between $\zeta$ and $\delta\sigma$ fields.}
\end{figure}

Since a negative mass leads to the growth of the field perturbations on superhorizon scales,
one may be curious about the contribution from higher-order corrections where more free fields are running in loops.
Up to quartic order in perturbations there are two kinds of non-derivative interactions $H_I^{(3)}\sim \epsilon a^4 V_{\sigma\sigma} \zeta \delta\sigma^2$ and 
$H_I^{(4)}\sim \epsilon^2 a^4 V_{\sigma\sigma} \zeta^2 \delta\sigma^2$ induced by the mass term,
as considered in the one-loop calculations. Each non-derivative interaction can introduce at most two free $\delta\sigma$ fields but inevitably
generates one more propagator $G_\zeta$ or $G_\sigma$. Higher-order non-derivative interactions with more $\zeta$ fields are further suppressed
by the slow-roll parameter since they must come from $\alpha = -\epsilon \zeta$ through the gravitational coupling.

It can be checked that if we want to insert a vertex of the quartic interaction $H_I^{(4)}$ 
as a 1PI diagram, then there are at most 4 free $\delta\sigma$ fields as the cases of Figure \ref{subfig-CIM_q} and \ref{subfig-CIM_cq}.
Further insertion of the quartic vertex as a 1PI diagram must introduce one internal propagator $G_\sigma$
(otherwise becomes non-1PI diagram if the internal propagator is $G_\zeta$). 
As a result the 1PI insertion of a quartic vertex does not change the time-dependence of the original diagram but only leads to an extra suppression factor
$\sim \epsilon M^2 H^2/M_p^2$, provided that $V_{\sigma\sigma} \ll M_p^2$.
For the cubic interaction $H_I^{(3)}$ we have to insert these vertices as a pair.
Similarly there are at most 4 free $\delta\sigma$ fields by inserting cubic vertices as of Figure~\ref{subfig-CIM_cq} and \ref{subfig-CIM_c}.
Further 1PI insertion of a cubic vertex cannot change the time-dependence but instead suppresses the original diagram by a factor
$\sim \epsilon M^4 H^2/M_p^2$ if we consider $\vert M^2\vert \ll \epsilon^{-1}\times 10^5$.
We therefore conclude that the one-loop $\rm CIM$ diagram (Figure \ref{subfig-CIM}) is the leading 1PI contribution to the $\zeta $ two-point function.

\begin{figure}
	\label{Fig_4}
	\begin{center}
			\includegraphics[width=6cm]{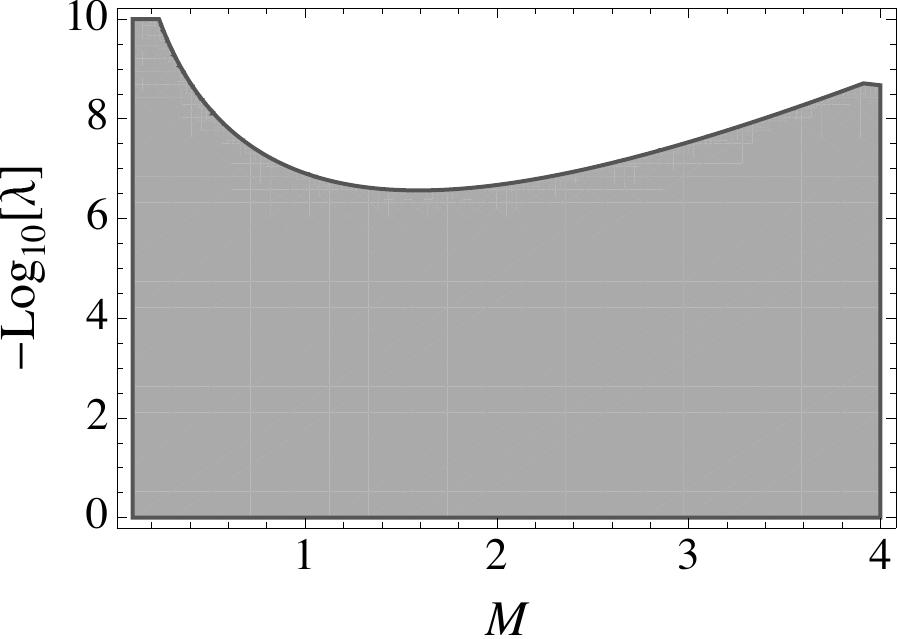}
	\end{center}
	\caption{\label{Fig lambda_M}The constraint of parameters with $\epsilon = 0.0068$ and $c_\ast = 1$. The white region is excluded by the perturbativity conditions.}
\end{figure}

The sum of the corrections to the $\zeta $ two-point function from the $\rm CIM$ diagram Figure~\ref{subfig-CIM} to all orders is schematically given by
\begin{align}
\label{eq:all_loop correction} 
\langle \zeta^2\rangle \sim & \frac{H^2}{\epsilon M_p^2}
\left[1+ c_\ast \frac{H^2}{\epsilon M_p^2}\epsilon^2 M^4\Pi_\zeta(a) + \left(c_\ast\frac{H^2}{\epsilon M_p^2} \epsilon^2 M^4\Pi_\zeta(a) \right)^2 + \dots \right], 
\end{align}
where $c_\ast$ is a constant factor of $\mathcal{O}(1-10)$, and  $\Pi_\zeta (a) \propto a^{4l-6}$ for $l \neq 3/2$, as given by \eqref{eq:Pi}.
Since $l = l (M) $, 
the time-dependent factor evaluated at a certain epoch $\Pi_\zeta(a_\ast) = \Pi_\zeta(M)$ is in fact a function of the mass.
In the case of a positive mass square ($l < 3/2$), higher-order terms decay rapidly so that loop corrections have no effects on the
physical observable. With a phase of a smooth transition of the field expectation value, 
there can be a period of $M^2 < 0$ ($l > 3/2$) where loop corrections grow with time
and the duration of the growing phase is also mass-dependent. In the scenario \eqref{eq:potential_sigma} with $g \rightarrow 0$, 
we always find a maximum correction
 around $\sqrt{\vert M^2\vert}= 2$ with a large enhancing factor $M^4 \Pi_\zeta (M)$ that depends on the model parameters. 
	We can constrain the value of $\lambda$ from the condition $M^4 \Pi_\zeta (M) < M_p^2/(c_\ast \epsilon H^2)$ for the breakdown of perturbative expansion.
Another perturbativity condition $\frac{\lambda}{4}v^4 \ll 3M_p^2 H^2$ given in Appendix \ref{Appendix} 
ensures that the massive field has a subdominant density during inflation.
The parameter space compatible with both two conditions is depicted in Fig. \ref{Fig lambda_M}.
For a large mass $\vert M^2  \vert =4$ we find $\lambda >  2\times 10^{-7}$ with $\epsilon = 0.0068$ from \eqref{eq:duration_of_slowroll}.

Before closing the discussion, we briefly comment the consequence due to the self-coupling of the massive field.
For example, a self-interaction $\mathcal{L}\sim a^4 \lambda \sigma^4$ as a part of the potential may allow us to perform
arbitrary 1PI insertion to any internal $\sigma$-field line with the introduction of two more free $\delta\sigma$ fields.
In this case the breakdown of perturbative expansion would request $\lambda < e^{\Delta N (3- 2l)} \simeq 9\times 10^{-7}$
with $\Delta N$ given by \eqref{eq:duration_of_slowroll} at $\sqrt{\vert M^2\vert}= 2$.
However the temporal divergence due to the self-coupling of a massive field is not specific for a negative mass square
and is also found in the case with a positive mass square \cite{Xue:2012wi}.
We believe that part of the temporal divergence should be removed by some renormalizations of the theory, 
as the finding from a resummation of the perturbative masses \cite{Weinberg:2006ac}.
A resummation of the quartic self-corrections of a massive field to the propagator in the $M^2 \ll 1$ limit
has been performed through the dynamical renormalization group method \cite{Burgess:2009bs,Burgess:2010dd},
and the question that how are the self-corrections behave after these treatments is left for future efforts.

\section{Conclusions}\label{sec:conclusion}

The transition of vacuum expectation values of scalar fields plays an essential role in cosmic inflation.
	In particular, many inflationary potentials naturally exhibit a concave (or convex upward) region 
	where the second derivative of the potential is  
	negative \cite{Starobinsky:1980te,Bezrukov:2007ep,Linde:1993cn,Brax:2010ai,Boubekeur:2005zm}.
In this work we have studied loop corrections to primordial fluctuations from a kind of phase transition realized in one of this type of potential, 
where a massless field randomly find its global minimum purely triggered by quantum fluctuations ($g=0$) or
smoothly rolls down into the minimum as an attractor ($g > 0$). 
The transitional phase involves with a classical evolution effectively driven by a negative mass term due to the convex shape of the potential,
 where field perturbations start to grow on superhorizon scales.

	We have applied the modified commutator form \cite{Senatore:2009cf,Musso:2006pt,Pimentel:2012tw} 
  to the loop computation, which is nothing but a rearrangement of the usual in-in formalism \cite{Weinberg:2005vy}.
The commutator form given by \eqref{eq:double-commutator} is most convenient for extracting the dominant contribution in the IR regime.
At one-loop level, there are three ``channels'' to generate primordial fluctuations in the bilinear correlation function.
We found that the most important corrections are generated by field perturbations that have been frozen outside the horizon
by the starting time of phase transition. The dominant diagram at this level is pratically a classical process where a second-order perturbation is created by
the scattering of two free-field perturbations. 
This result is opposite to the usual case with a positive field mass, where the dominant channel shall have the least number of free fields.
We also found that the resulting loop corrections are only sensitive to those field masses comparable to
the Hubble scale of inflation. 
However, such a kind of phase transition cannot generate significant tensor mode perturbations at second order, given that the leading interaction
involved with the graviton coupling must contain spatial derivatives. 

	We emphasize that the above conclusions are not changed by the UV completion of the theory, 
since the dominant corrections from non-derivative interactions led by \eqref{eq:CIM_result} are generated around the IR cutoff. 
To study the UV physics, it is more convenient to use different arrangement of the in-in formalism \cite{Chen:2010xka},
but the UV effects are independent of the spectator field potentials and thus are expected to have only subdominant importance on superhorizon scales.
For instance, the Fadeev-Popov ghost due to the general covariance of the theory \cite{Armendariz-Picon:2014xda} 
is always subject to derivative interactions and therefore its contributions cannot be large on scales of our interest.

Based on the above findings we can gain some knowledge of loop effects induced by other kinds of phase transitions.
For example a time-dependent transition of the global minimum similar to that of the Higgs mechanism has been considered in \cite{Nagasawa:1991zr}.
In this kind of models there also exists a period of negative-mass evolution, yet no field perturbations can survive on superhorizon scales by the time of the 
growing phase. As a result loop corrections are only generated by those modes that exit the horizon during the negative-mass evolution, which
effectively invokes a scale-dependence in the calculation and thus suppresses the final contribution.

Though the specific scenario investigated in this paper only realizes a large enough correction to the observables in a certain parameter range, 
we find our results interesting because in reality the physics relevant to inflation should be much more complicated.
	We remark that our results based on only the specific time-interval $\mathcal{I}_2$ in fact underestimates the real
	contributions from the potential \eqref{eq:potential_sigma} (or from 
	the well-known potentials \cite{Boubekeur:2005zm,Lyth:1998xn,Brax:2010ai,Starobinsky:1980te,Bezrukov:2007ep,Linde:1993cn}), 
given that the initial values of $\mathcal{I}_3$ are also enhanced by the growing phase, 
and the actual effect may be even larger than what we have shown here.
It is unclear whether there could be a more efficient way for spectator fields to affect primordial fluctuations during inflation, 
but the current conclusion indicates that loop corrections residing in the observed cosmological correlations may
be more important than they are usually expected.

\acknowledgments

We thank Robert Brandenberger, Xian Gao, Takahiro Hayashinaka, Minxi He, Tomohiro Nakama, Yuki Watanabe, Yi Wang and Siyi Zhou for helpful discussions.
We are in particular grateful to David Seery and Teruaki Suyama for many useful comments.
Y. P. W. was supported by Ministry of Science and Technology (MoST) Postdoctoral Research Abroad Program (PRAP)
MoST-105-2917-I-564-022, 
and is currently supported by JSPS International Research Fellows and JSPS KAKENHI Grant-in-Aid for Scientific Research No. 17F17322.
J. Y. was supported by JSPS
KAKENHI, Grant-in-Aid for Scientific Research
No. 15H02082, and Grant-in-Aid for Scientific Research
on Innovative Areas No. 15H05888.

\appendix
\section{Linear perturbation equations}\label{Appendix}

We solve the field equations by using the metric \eqref{eq:metric_eta} and the decomposition 
$\phi(\mathbf{x}, \eta) =\phi(\eta) + \delta\phi(\mathbf{x}, \eta)$ and 
$\sigma(\mathbf{x},\eta) =\sigma(\eta) + \delta\sigma(\mathbf{x}, \eta)$.
The equation of motion for the homogeneous parts are
\begin{align}
3\mathcal{H}^2 &=\frac{\kappa^2}{2}(\phi^{\prime 2}+\sigma^{\prime 2}) +a^2 \kappa^2V(\phi,\sigma), \\
\mathcal{H}^2 + 2\mathcal{H}^\prime &= -\frac{\kappa^2}{2}(\phi^{\prime 2}+\sigma^{\prime 2}) + a^2 \kappa^2V(\phi,\sigma), \\
0&= \phi^{\prime\prime} + 2\mathcal{H}\phi^\prime + V_\phi ,\\
\label{eq:eom_sigma_classical}
0&= \sigma^{\prime\prime} + 2\mathcal{H}\sigma^\prime + V_\sigma .
\end{align}
At linear order, the relevant Einstein equations for scalar perturbations are 
\begin{align}
6\mathcal{H}^2\alpha +2\mathcal{H}\partial^2\beta &= \kappa^2(\phi^{\prime 2}\alpha-\phi^\prime\delta\phi^\prime -a^2V_\phi\delta\phi)
+ \kappa^2(\sigma^{\prime 2}\alpha-\sigma^\prime\delta\sigma^\prime -a^2V_\sigma\delta\sigma), \\
2\mathcal{H}\alpha &= \kappa^2(\phi^\prime\delta\phi +\sigma^\prime\delta\sigma), \\
0 &= \alpha + \beta^\prime + 2 \mathcal{H}\beta  ,
\end{align}
and the field perturbations follow
\begin{align}
\label{eom:delta_phi_full}
\delta\phi^{\prime\prime} + 2\mathcal{H}\delta\phi^\prime -\partial^2\delta\phi +a^2V_{\phi\phi}\delta\phi + 
2( \phi^{\prime\prime} + 2\mathcal{H}\phi^\prime )\alpha + \phi^\prime(\alpha^\prime +\partial^2\beta) &=0, \\
\label{eom:delta_sigma_full}
\delta\sigma^{\prime\prime} + 2\mathcal{H}\delta\sigma^\prime -\partial^2\delta\sigma+a^2V_{\sigma\sigma}\delta\sigma + 
2( \sigma^{\prime\prime} + 2\mathcal{H}\sigma^\prime )\alpha + \sigma^\prime(\alpha^\prime +\partial^2\beta) &=0.
\end{align}
By using the above equations we can derive
\begin{align}
\label{sol:alpha_1}
\alpha &= \kappa^2 \left(\frac{\phi^\prime}{2\mathcal{H}}\delta\phi +\frac{\sigma^\prime}{2\mathcal{H}}\delta\sigma \right), \\
\label{sol:beta_1}
\partial^2\beta &= \frac{\kappa^2\phi^{\prime 2}}{2\mathcal{H}^2} \left(-\frac{\mathcal{H}}{\phi^\prime}\delta\phi\right)^\prime +
\frac{\kappa^2\sigma^{\prime 2}}{2\mathcal{H}^2} \left(-\frac{\mathcal{H}}{\sigma^\prime}\delta\sigma\right)^\prime .
\end{align}
We are interested in the case $V(\phi, \sigma) =V_0(\phi) + V_1(\sigma) = V_0(1+f)$, 
where $f = V_1/V_0$ is the fraction of the potential energy of $\sigma$and $\phi$. 
Therefore it is convenient to define $\epsilon_\phi = \kappa^2\phi^{\prime \, 2}/(2 \mathcal{H}^2)$,
$\epsilon_\sigma = \kappa^2\sigma^{\prime 2}/(2a^2 V_1)$, $\zeta_\phi = -\mathcal{H}\delta\phi/\phi^\prime$,
$\zeta_\sigma = -\mathcal{H}\delta\sigma/\sigma^\prime$, such that \eqref{sol:alpha_1} and \eqref{sol:beta_1} become
\begin{align}
\label{sol:alpha_2}
\alpha &= -\epsilon_\phi \zeta_\phi - \frac{3f}{1+f} \epsilon_\sigma \zeta_\sigma , \\
\label{sol:beta_2}
\partial^2\beta &= \epsilon_\phi \zeta_\phi^\prime + \frac{3f}{1+f} \epsilon_\sigma \zeta_\sigma^\prime.
\end{align}
	Then the result of the single field inflation is recovered in the limit $f\rightarrow 0 $ even with $\epsilon_\sigma \sim \mathcal{O}(1)$.
Since the potential energy released from the phase transition of \eqref{eq:potential_sigma} is $\lambda v^4/4$, the condition $f \ll 1$ implies $\lambda v^4 \ll 12 M_p^2 H^2$. In terms of the mass parameter $M_i^2 = \lambda v^2/H^2$, we find the condition $v^2 < 12 M_p^2/M_i^2$.

\section{Second-order perturbation equations}\label{Appendix B}

Here we examine the mass-induced quartic interactions by solving some of the field equations up to second order.
We first parametrize the perturbations up to second order (ignoring all vector modes) as
\begin{align}\label{eq:2nd parametrization}
N & = 1+\alpha +\alpha_2, \nonumber\\
\beta^i &= \partial_i \beta + \partial_i \beta_2, \\ \nonumber
\gamma_{ij} &= a^2\left[1+ h^{(2)}_{ij}\right], 
\end{align}
where we will not consider the contribution from the first-order tensor mode. For loop corrections from the graviton at 
first order one may refer to \cite{Dimastrogiovanni:2008af}.

At second order, the Einstein equation for the momentum flux and the anisotropy stress are given respectively by 
\begin{align}\nonumber \label{eq:2nd momentumflux}
& 6\mathcal{H}\alpha\partial_i\alpha  + \partial_i\alpha \partial^2\beta - \partial_i\partial_j\beta \partial^j\alpha
 -2\mathcal{H}  \partial_i\alpha_2= \\
&\qquad\qquad-\kappa^2 (\delta\phi^\prime\partial_i\delta\phi + \delta\sigma^\prime\partial_i\delta\sigma- 2\phi^\prime \alpha \partial_i\delta\phi
- 2\sigma^\prime \alpha \partial_i\delta\sigma) ,\\ 
\nonumber 
\label{eq:2nd anisotropy}
&\frac{1}{4}(h_{ij}^{\prime\prime}+2 \mathcal{H}h_{ij}^\prime  -\partial^2 h_{ij}) 
+ \partial^k\beta\partial_k(\partial_i\partial_j\beta) +\partial^2\beta\partial_i\partial_j\beta 
-\alpha\partial_i \partial_j \alpha + \alpha^\prime \partial_i \partial_j\beta  \\
&\qquad\qquad 
-\partial_i\partial_j\alpha_2 -\partial_i\partial_j (\beta_2^\prime + 2\mathcal{H}\beta_2) =
\kappa^2(\partial_i \delta\phi \partial_j\delta\phi + \partial_i\delta\sigma \partial_j \delta\sigma),
\end{align}
where $h_{ij}\equiv h^{(2)}_{ij}$ unless otherwise mentioned.
According to \eqref{eq:2nd anisotropy}, at fourth-order in the perturbative expansion 
there is only one interaction with a pair of $\delta\sigma$ fields and a tensor mode as given by \eqref{eq:interaction_hss}.

By using solutions of the linear perturbations, we can solve $\alpha_2$ from \eqref{eq:2nd momentumflux} as
\begin{align} \label{sol:alpha2}
\nonumber
\alpha_2 = \frac{\alpha^2}{2} 
+& \frac{1}{2\mathcal{H}}\partial^{-2}\left[\partial^i( \partial_i\alpha \partial^2\beta - \partial_i\partial_j\beta \partial^j\alpha)\right] \\
&\qquad\qquad +  
\frac{\kappa^2}{2\mathcal{H}}\partial^{-2}\left[\partial^i(\delta\phi^\prime\partial_i\delta\phi + \delta\sigma^\prime\partial_i\delta\sigma)\right].
\end{align}
One can check that at fourth-order in the perturbative expansion, the mass-induced interaction
is only led by $\mathcal{L}^{(4)}\supset a^4 V_{\sigma\sigma} \delta\sigma^2\alpha_2 $.
Therefore, we find from \eqref{sol:alpha2} that the only non-derivative interaction of our interest is given by
\begin{align}
\mathcal{L}^{(4)}_{\rm nd} = \frac{a^4}{2}V_{\sigma\sigma}\alpha^2\delta\sigma^2 .
\end{align}

\section{General solutions of the mode function}\label{Appendix C}

The general solution of the coefficinets $b_1$ and $b_2$ in the mode function \eqref{sol:sigma_k phase2} with initial values $\delta\sigma_k(z_i)$ and $\delta\sigma^\prime_k(z_i)$
given by \eqref{sol:sigma_k phase1} reads
\begin{align}
b_1 &= -\frac{\pi^{3/2}}{4}\left[z_i Y_{l_-}(-z_i)H^{(1)}_{1/2}(-z_i)+(-z_i Y_{l_{-}-1}(-z_i)+(\frac{3}{2}-l_{-})Y_{l_-}(-z_i))H^{(1)}_{3/2}(-z_i)\right], \\
b_2 &= i\frac{\pi^{3/2}}{4}\left[-z_i J_{l_-}(-z_i)H^{(1)}_{1/2}(-z_i)+(z_iJ_{l_- -1}(-z_i)-(\frac{3}{2}-l_{-})J_{l_-}(-z_i))H^{(1)}_{3/2}(-z_i)\right].
\end{align}
Here $H^{(1)}_l$ is Hankel function of the first kind. The mode function \eqref{sol:sigma_k phase2} is thus of the form
\begin{align}\label{eq:general sigma_k phase2}
\delta\sigma_k =& \frac{H}{\sqrt{2k^3}}\frac{\pi}{4}e^{-i z_i}\left(\frac{z}{z_i}\right)^{3/2} 
\left[B_1(l_-,z_i)J_{l_-}(-z) +B_2(l_-,z_i)Y_{l_-}(-z)\right],
\end{align}
where the coefficients are
\begin{align}
B_1 &= 2z_i(z_i-i)Y_{l_- -1}(-z_i)  +[3i - (3+2iz_i)z_i +2l_{-}(z_i-i)]Y_{l_-}(-z_i),
\\
B_2 &= -2z_i(z_i-i)J_{l_- -1}(-z_i)+[-3i +(3+2iz_i)z_i-2l_{-}(z_i-i)]J_{l_-}(-z_i).
\end{align}
In the late-time limit where $z\rightarrow 0$, the solution \eqref{eq:general sigma_k phase2} is led by the terms
\begin{align}
	\delta\sigma_k = \frac{H}{\sqrt{2k^3}}\frac{\pi}{4}e^{-i z_i} \left(\frac{z}{z_i}\right)^{3/2} 
	\left[\frac{B_1}{\Gamma(l+1)}\left(\frac{-z}{2}\right)^l - \frac{B_2 \,\Gamma(l)}{\pi}\left(\frac{-z}{2}\right)^{-l}+ \cdots\right],
\end{align}
where $l = l_-$. If a $\delta\sigma$ field is produced during this phase, 
the leading terms of the Green function \eqref{eq:Green_sigma} in the late-time limit is then given by
\begin{align}\label{eq:Green_sigma I2}
	G_{\bf k}^{\sigma} (z ; \tilde{z}) 
	= i \theta (z - \tilde{z}) \frac{\pi H^2}{32 k^3} \frac{\Gamma(l)}{\Gamma(l+1)}
    \left(\frac{z \tilde{z}}{z_i^2}\right)^{3/2} \left[\left(\frac{\tilde{z}}{z}\right)^l - \left(\frac{z}{\tilde{z}}\right)^l\right]
    \left(B_1 B_2^\ast +B_2 B_1^\ast\right).
\end{align}

Similarly, the general solution of the coefficinets $c_1$ and $c_2$ in the mode function \eqref{sol:sigma_k phase3} 
with initial values $\delta\sigma_k(z_f)$ and $\delta\sigma^\prime_k(z_f)$
given by \eqref{sol:sigma_k phase2} reads
\begin{align}
c_1 =& -\frac{\pi}{2}\left\lbrace z_f Y_{l_+ -1}(-z_f) 
\left[b_1 J_{l_-}(-z_f)+ i b_2 Y_{l_{-}}(-z_f)\right] \right.\\\nonumber
&+Y_{l_+ }(-z_f)[b_1 z_f J_{l_- -1}(-z_f) +ib_2 z_f Y_{l_- -1}(-z_f) \\\nonumber
&\left.+(l_- - l_+)(b_1  J_{l_- -1}(-z_f)+ib_2  Y_{l_- -1}(-z_f))]\right\rbrace, \\
c_2 =&  -\frac{\pi}{2}\left\lbrace z_f J_{l_+ }(-z_f) 
\left[-i b_1 J_{l_- -1}(-z_f)+ i b_2 Y_{l_{-}-1}(-z_f)\right] \right.\\\nonumber
&+i b_1 J_{l_- }(-z_f)[z_fb_1  J_{l_+ -1}(-z_f) -(l_- - l_+) J_{l_+ }(-z_f)] \\\nonumber
&\left.+b_2Y_{l_-}(-z_f)[-z_f  J_{l_+ -1}(-z_f)+(l_- -l_+)  J_{l_+ }(-z_f)]\right\rbrace.
\end{align}
Since $b_1$ and $b_2$ only depend on $l_-$ and $z_i$, the mode function \eqref{sol:sigma_k phase3} in fact takes the form
\begin{align}
\delta\sigma_k =& \frac{H}{\sqrt{k^3}}\frac{\pi}{2}\left(-z \right)^{3/2} 
\left[C_1(l_-,l_+,z_i,z_f)J_{l_+}(-z) +C_2(l_-,l_+,z_i,z_f)Y_{l_+}(-z)\right],
\end{align}
where the coefficients are
\begin{align}
C_1 =& z_fY_{l_+ -1}(-z_f)[b_1 J_{l_-}(-z_f)+ i b_2 Y_{l_-}(-z_f)]  \\\nonumber
&+Y_{l_+}(-z_f)[b_1 z_f J_{l_- -1}(-z_f)+ i b_2 z_f Y_{l_- -1}(-z_f)  \\\nonumber
&+ (l_{-} - l_+) (b_1 J_{l_-}(-z_f)+ i b_2 Y_{l_-}(-z_f))],
\\
C_2 =& - z_fJ_{l_+ }(-z_f)[ b_1 J_{l_- -1}(-z_f)+ i b_2 Y_{l_- -1}(-z_f)] \\\nonumber
&+b_1J_{l_-}(-z_f)[z_f J_{l_+ -1}(-z_f) -(l_{-} - l_+) J_{l_+}(-z_f)] \\\nonumber
&-i b_2Y_{l_-}(-z_f)[-z_f J_{l_+ -1}(-z_f) +(l_{-} - l_+)J_{l_+}(-z_f)].
\end{align}
One can use this solution to compute the Green function $G_\sigma$ for a $\delta\sigma$ field generated during this phase.

\section{$\mathcal{I}_2$ contribution of the one-loop {\rm CIS} correlation functions}\label{Appendix D}
At one-loop level there are at most two time-integrations in each diagram, and each time-intgration is divided into three parts as \eqref{def:t-integration split}. 
We compute the pure $\mathcal{I}_2$ contribution in the one-loop correlation $\langle\zeta^2\rangle_{\rm CIS,1}$
with the cubic interaction given by $\mathcal{H}_I^{(3)}= \epsilon_\phi \frac{a^4}{2}\lambda v^2\zeta\delta\sigma^2$.
In the momentum space, this two-point function reads
\begin{align}
 \langle \zeta_\mathbf{k}(\eta)\zeta_\mathbf{K}(\eta)& \rangle_{\rm CIS,1} \supset  \;
 8 (\lambda v^2)^2 \delta({\bf k +K}) \;
 \mathrm{Re} \left[\int_{\eta_i}^{\eta_f}d\eta_2\int_{\eta_i}^{\eta_f}d\eta_1 \epsilon_2a^4(\eta_2)\epsilon_1a^4(\eta_1) \right. \\\nonumber
 \times& \left.  \int \frac{d^3 \mathbf{p}_1}{(2\pi)^3}\frac{d^3 \mathbf{q}_1}{(2\pi)^3}\delta({\bf k-p_1+q_1}) 
 G_{\bf p_1}^\sigma(\eta_2;\eta_1 )G^\zeta_{\bf k}(\eta;\eta_2) 
 \zeta_k(\eta_1)\zeta_k^\ast(\eta) \delta\sigma_{q_1 }(\eta_1)\delta_{q_1}^\ast(\eta_2)\right].
\end{align}
We are interest in internal modes (such as $p_1$ and $q_1$) that exit the horizon by $\eta_i$ where the mode function
\eqref{sol:sigma_k small mass} can be used. The $\zeta$ Green function is given by \eqref{eq:Green_zeta} and 
the $\sigma$ Green function in the late-time limit is given in appendix \ref{Appendix C}.
For $k < k_i$, we find that
\begin{align}\label{eq:co_CIS1 k<ki}
	 \langle \zeta_\mathbf{k}(\eta)\zeta_\mathbf{K}(\eta) \rangle_{\rm CIS,1} &\supset  \;
	 \frac{\kappa^4 H^4}{8 k^3}M_-^4 \delta({\bf k + K}) \; \mathrm{Re} \left[\mathcal{I}_{\rm CIS,1}\right] \\\nonumber
	 \times & \frac{2}{3z_i^3} \left\vert\int \frac{dz_2}{z_2^4}\theta(z-z_2)(z^3-z_2^3)\left(\frac{z_2}{z_i}\right)^{3-2l}
	 \left(\frac{1}{2l}+\ln\left(\frac{z_2}{z_i}\right)\right) \right\vert,
\end{align} 
where we define the mode-integration from \eqref{eq:Green_sigma I2} as
\begin{align}
	\mathcal{I}_{\rm CIS,1} \equiv \frac{1}{k^3}  \int \frac{d^3 \mathbf{p}_1}{(2\pi)^3}\frac{d^3 \mathbf{q}_1}{(2\pi)^3}
	\delta({\bf k-p_1+q_1}) \frac{\pi \Gamma(l)}{16 \Gamma(l +1)} \frac{1}{p_1^3 q_1^3}(B_1B_2^\ast + B_2 B_1^\ast).
\end{align}
The leading $\mathcal{I}_2$ contribution in \eqref{eq:co_CIS1 k<ki} then reads
\begin{align}
	\langle \zeta_\mathbf{k}(\eta)\zeta_\mathbf{K}(\eta) \rangle_{\rm CIS,1} \supset  \;
	\frac{\kappa^4 H^4}{8 k^3}M_-^4 \delta({\bf k + K}) \; \mathrm{Re} \left[\mathcal{I}_{\rm CIS,1}\right] 
	\frac{2}{3z_i^3}\frac{\ln (z_f/z_i)}{l(2l-3)}\left(\frac{z_f}{z_i}\right)^{3-2l}.
\end{align}

We now move to the $\mathcal{I}_2$ contribution in the one-loop correlation $\langle\zeta^2\rangle_{\rm CIS,2}$
with the quartic interaction given by $\mathcal{H}_I^{(4)}= \epsilon_\phi^2 \frac{a^4}{2}\lambda v^2\zeta^2\delta\sigma^2$.
In the momentum space,
\begin{align}
	 \langle \zeta_\mathbf{k}(\eta)\zeta_\mathbf{K}(\eta) \rangle_{\rm CIS,2} \supset & \;
	 2\lambda v^2  \delta({\bf k +K})  \;
	 \mathrm{Im}\left[\int_{\eta_i}^{\eta_f}d\eta_1 \epsilon_1^2 a^4(\eta_1) G^\zeta_{\bf k}(\eta;\eta_1) \right.\\\nonumber
	  &\times \left.  \zeta_k(\eta_1)\zeta_k^\ast(\eta) \int \frac{d^3 \mathbf{p}_1}{(2\pi)^3}
	  \delta\sigma_{p_1 }(\eta_1)\delta_{p_1}^\ast(\eta_1)\right].
\end{align}
For $k < k_i$ and $p_1 < k_i$, we use \eqref{sol:sigma_k small mass} and \eqref{eq:Green_zeta} to derive
\begin{align}
	\langle \zeta_\mathbf{k}(\eta)\zeta_\mathbf{K}(\eta) \rangle_{\rm CIS,2} \supset  \;
	\frac{\kappa^4H^4}{8\pi k^3 } M_-^2  \delta({\bf k +K})  \; \mathrm{Im}
	\left[\frac{i}{2l(3-2l)}\left(\frac{z_f}{z_i}\right)^{3-2l}\ln\left(\frac{k}{k_0}\right)\right],
\end{align}
where the integration function \eqref{def:F_n} has been used.


\end{document}